\documentclass[11pt]{article}

\usepackage[letterpaper,margin=0.98in]{geometry}
\usepackage{amsmath, amssymb, amsthm, thmtools, amsfonts}
\usepackage{bbm}

\usepackage{ifthen}

\usepackage{cite}
\usepackage{appendix}
\usepackage{graphicx}
\usepackage{color}
\usepackage{algorithm}
\usepackage[noend]{algpseudocode}
\usepackage{epstopdf}
\usepackage[textsize=tiny]{todonotes}

\usepackage{framed}
\usepackage[framemethod=tikz]{mdframed}
\usepackage[bottom]{footmisc}
\usepackage[shortlabels]{enumitem}
\setitemize{noitemsep,topsep=3pt,parsep=3pt,partopsep=3pt}
\usepackage[font=small]{caption}
\usepackage{xspace}

\definecolor{darkgreen}{rgb}{0,0.5,0}
\usepackage{hyperref}
\hypersetup{
    unicode=false,          
    colorlinks=true,        
    linkcolor=red,          
    citecolor=darkgreen,        
    filecolor=magenta,      
    urlcolor=cyan           
}
\usepackage[capitalize, nameinlink]{cleveref}
\crefname{theorem}{Theorem}{Theorems}
\Crefname{lemma}{Lemma}{Lemmas}

\newcommand{\f}{\displaystyle\frac}
\newcommand{\cd}{\cdot}

\newcommand{\sr}{\sqrt}

\newcommand{\lds}{\ldots}

\newcommand{\s}{\subseteq}

\newcommand{\BE}{\begin{enumerate}}
\newcommand{\EE}{\end{enumerate}}
\newcommand{\im}{\item}
\newcommand{\BI}{\begin{itemize}}
\newcommand{\EI}{\end{itemize}}

\newcommand{\logn}{\log n}

\newcommand{\N}{\mathbb N}

\newcommand{\eps}{\epsilon}
\newcommand{\e}{\epsilon}
\newcommand{\de}{\delta}

\newcommand{\al}{\alpha}

\newcommand{\Om}{\Omega}
\newcommand{\el}{\ell}

\newcommand{\Th}{\Theta}

\newcommand{\tO}{\tilde{O}}

\newcommand{\lf}{\lfloor}
\newcommand{\rf}{\rfloor}
\newcommand{\lc}{\lceil}
\newcommand{\rc}{\rceil}

\newcommand{\poly}{\text{poly}}

\newcommand{\lp}{\left(}
\newcommand{\rp}{\right)}
\newcommand{\lb}{\left[}
\newcommand{\rb}{\right]}
\newcommand{\lmt}{\left[\begin{matrix}}
\newcommand{\rmt}{\end{matrix}\right]}

\newtheorem{theorem}{Theorem}[section]
\newtheorem{lemma}[theorem]{Lemma}
\newtheorem{corollary}[theorem]{Corollary}
\newtheorem{claim}[theorem]{Claim}
\newtheorem{observation}[theorem]{Observation}
\newtheorem{fact}[theorem]{Fact}
\newtheorem{remark}[theorem]{Remark}
\newcommand{\BT}{\begin{theorem}}
\newcommand{\ET}{\end{theorem}}
\newcommand{\BL}{\begin{lemma}}
\newcommand{\EL}{\end{lemma}}
\newcommand{\BC}{\begin{corollary}}
\newcommand{\EC}{\end{corollary}}
\newcommand{\BCL}{\begin{claim}}
\newcommand{\ECL}{\end{claim}}
\newcommand{\BO}{\begin{observation}}
\newcommand{\EO}{\end{observation}}
\newcommand{\BF}{\begin{fact}}
\newcommand{\EF}{\end{fact}}
\newcommand{\BP}{\begin{proof}}
\newcommand{\EP}{\end{proof}}
\newcommand{\BPS}{\begin{proof}[Proof (Sketch)]}
\newcommand{\EPS}{\end{proof}}

\algnewcommand{\IIf}[1]{\State\algorithmicif\ #1\ \algorithmicthen}
\algnewcommand{\EndIIf}{\unskip\ \algorithmicend\ \algorithmicif}
\algrenewcommand\algorithmiccomment[2][\normalsize]{{#1\hfill\(\triangleright\) \emph{#2}}}

\newcommand{\defn}[1]{{\emph{#1}}}

\newcommand{\congest}{$\mathsf{CONGEST}$\xspace}

\makeatletter
\newcounter{algocounter}
\@ifpackageloaded{hyperref}%
  {\newcommand{\mylabel}[2]
    {\refstepcounter{algocounter}\protected@write\@auxout{}{\string\newlabel{#1}{{\textcolor{blue}{\textup{#2}}}{\thepage}%
      {\@currentlabelname}{\@currentHref}{}}}}}%
\makeatother

\newcommand{\FullOrShort}{short}

\ifthenelse{\equal{\FullOrShort}{full}}{
        
  \newcommand{\fullOnly}[1]{#1}
  \newcommand{\shortOnly}[1]{}
  
  }{

    \newcommand{\fullOnly}[1]{}
  }

\begin{document}

\date{}

\title{Improved Distributed Algorithms for Exact Shortest Paths}

\author{
        Mohsen Ghaffari\\
  \small ETH Zurich \\
  \small ghaffari@inf.ethz.ch
        \and
        Jason Li\\
  \small CMU \\
  \small jmli@cs.cmu.edu
 }
\maketitle

\begin{abstract}
Computing shortest paths is one of the central problems in the theory of distributed computing. For the last few years, substantial progress has been made on the \textit{approximate} single source shortest paths problem, culminating in an algorithm of \textcolor{black}{Becker et al.~[DISC'17]} which deterministically computes $(1+o(1))$-approximate shortest paths in $\tilde O(D+\sqrt n)$ time, where $D$ is the hop-diameter of the graph. Up to logarithmic factors, this time complexity is optimal, matching the lower bound of Elkin~[STOC'04].

\medskip
The question of \textit{exact} shortest paths however saw no algorithmic progress for decades, until the recent breakthrough of Elkin [STOC'17], which established a sublinear-time algorithm for exact single source shortest paths on undirected graphs. Shortly after, Huang et al.\ [FOCS'17] provided improved algorithms for exact all pairs shortest paths problem on directed graphs.
\medskip

In this paper, we present a new single-source shortest path algorithm with complexity $\tilde O(n^{3/4}D^{1/4})$. For polylogarithmic $D$, this improves on Elkin's $\tilde{O}(n^{5/6})$ bound and gets closer to the $\tilde{\Omega}(n^{1/2})$ lower bound of Elkin~[STOC'04].  For larger values of $D$, we present an improved variant of our algorithm which achieves complexity $\tO\lp n^{3/4+o(1)}+ \min\{ n^{3/4}D^{1/6},n^{6/7}\}+D\rp$, and thus compares favorably with Elkin's bound of $\tilde{O}(n^{5/6} + n^{2/3}D^{1/3} + D ) $ in essentially the entire range of parameters. This algorithm provides also a qualitative improvement, because it works for the more challenging case of directed graphs (i.e., graphs where the two directions of an edge can have different weights), constituting the first sublinear-time algorithm for directed graphs. Our algorithm also extends to the case of exact $\kappa$-source shortest paths, giving a complexity of $\tO \lp \min \left\{  \kappa^{1/2}n^{3/4+o(1)} + \kappa^{1/3}n^{3/4}D^{1/6}  , \kappa^{3/7}n^{6/7} \right\} + D \rp$. For moderately small $\kappa$ and $D$, this improves on the $\tilde{O}({\kappa}^{1/2}n^{3/4}+n)$ bound of Huang et al.
\end{abstract}

\setcounter{page}{0}
\thispagestyle{empty}
\newpage

\section{Introduction \& Related Work}
\vspace{-0.3cm}
Computing shortest paths---either from a single node to all or from multiple or even all nodes to all---are among the most central problems in the theory of distributed computing, with consequential applications; e.g., they are a center-piece of routing in computer networks. As such, shortest path problems and distributed algorithms for them have been studied extensively since the 1950's, starting with the algorithms of Bellman\cite{bellman1958routing} and Ford\cite{ford1956network}. In this paper, we present new algorithms that improve on the state of the art of this decades-old problem.  Next, we first overview the background of the problem and the known results, and then we outline our contributions.

\vspace{-0.2cm}
\subsection{Background, Related Work, and State of the Art}
\vspace{-0.2cm}
\paragraph{Model} We work with the standard synchronous message-passing model of distributed computing, called \congest\cite{Peleg:2000}, where the network is abstracted as an undirected graph $G=(V,E)$ with one node (processor) at each vertex in $V$. Each edge has an integer weight in range $[1, \Lambda]$; we assume that the weights can be described in $\Theta(\log n)$ bits and thus $\Lambda\leq \poly(n)$. In general, we allow the two directions of an edge to have asymmetric weights, i.e., the weight of the link $v\rightarrow u$ might be different than that of $u\rightarrow v$. At the beginning, each node $v$ knows only its neighbors and the weights of the corresponding incident edges. The communication network is independent of the weights and is thus bi-directional. Per round, each node $u\in V$ can send an $O(\log n)$-bit message to each node $v\in V$ iff $\{u,v\}\in E$. Moreover, each node can perform arbitrary computations given the information that it has at that point. 

\vspace{-0.15cm}
\paragraph{The SSSP Problem} The \defn{single-source shortest path (SSSP)} problem is to compute $d(s,t)$ for each $t\in V$, given a source vertex $s\in V$. In the distributed setting, each node $t\in V$ should learn its distance $d(s,t)$ from the source. 

\vspace{-0.15cm}
\paragraph{State of the Art For Exact SSSP}
The classic approach of Bellman-Ford\cite{bellman1958routing, ford1956network} gives an $O(n)$ round algorithm for SSSP. This complexity is optimal in the worst case, in the sense that in graphs of hop-diameter $D=\Theta(n)$---e.g., think of a cycle---one cannot do better. But this is an unfortunate and uninformative impossibility. As was argued by Garay, Kutten, and Peleg in their influential work~\cite{Garay-Kutten-Peleg}, the case of graphs with small diameter $D$ is far more relevant in the real world, and far more interesting theoretically, and the above impossibility does not preclude faster algorithms for those cases. Since then, the case of network graphs with moderately low-diameter has become the focus point of the area of distributed graph algorithms, e.g., see the progress on problems such as minimum spanning tree\cite{Garay-Kutten-Peleg, Kutten-Peleg, DasSarma-11, Elkin-2004, Peleg-Rubinovich-1999, Planar-ShortCut}, min-cut\cite{distributed-cut,nanongkai2014cut}, etc. Howeover, despite this change of focus, for a long time, there was no progress on the algorithmic side of exact SSSP. The only development was an influential work of Elkin~\cite{Elkin-2004}, which proved that computing the SSSP requires $\Omega(n^{1/2})$ rounds, even on graphs with diameter as low as $D=O(\log n)$.

\vspace{-0.2cm}
\paragraph{A Turn Towards Approximation Algorithms} Due to the lack of progress on the algorithmic side of the exact SSSP, the area turned focus to the approximate version. It can be argued that this was sparked by the visionary article of Elkin\cite{elkin2004distributed} that emphasized approximations. Over the years, this led to a remarkable sequence of work\cite{DasSarma-11, Lenzen-PattShamir, Danupon-paths, henzinger2016deterministic, becker2016approximate}, providing better and better distributed SSSP approximation algorithms. By now, we can say that we have a good understanding of the approximate version of the problem, having the deterministic algorithm of \textcolor{black}{Becker et al.\cite{becker2016approximate}, which computes a $(1+o(1))$-approximation\footnote{Their algorithm computes a $(1+\e)$-approximate SSSP in $\tO(\e^{-O(1)}(\sr n+D))$ rounds, so we can take, e.g., $\e:=1/\log n$.} of SSSP in $\tilde{O}(D+\sqrt{n})$ rounds}, and whose complexity is known to be the best possible up to logarithmic factors, due to a lower bound of Elkin~\cite{Elkin-2004}. More precisely, the latter exhibits graphs of diameter $D=O(\log n)$ in which no SSSP approximation algorithm---even for rather coarse approximations---can run in less than $\tilde{\Omega}(\sqrt{n})$ rounds.

\paragraph{Return of the Exact Algorithms} Despite the numerous steps of progress that brought us to an almost complete understanding of the approximate case of SSSP, computing exact SSSP remained open until very recently. But then came the breakthrough of Elkin\cite{elkin2017SSSP}, which provided the first sublinear-time algorithm for exact SSSP. More precisely, Elkin gave an algorithm that computes exact SSSP in $\tilde{O}(n^{5/6})$ rounds if $D=\tilde{O}(\sqrt{n})$, and in $\tilde{O}(D^{1/3} n^{2/3})$ rounds if $D=\tilde{\Omega}(\sqrt{n})$. 

Shortly after, another significant step of progress was made on the closely related problem of All-Pairs Shortest Path (APSP), when Huang et al.\cite{huang2017distributed} gave an algorithm with complexity $O(n^{5/4})$ rounds for positive integer-weighted graphs\footnote{Elkin pointed out shortly after the work of Huang et al.\cite{huang2017distributed}, but independently from them, that the techniques in his work\cite{elkin2017SSSP} immediately lead to an exact APSP algorithm with complexity $\tilde{O}(n^{5/3})$. It is also worth noting that Elkin's algorithm works for nonnegative real weights, not just integer weights.}. Similar to SSSP, also for APSP, the $(1+o(1))$-approximate version of the problem is well-understood by now, having $\tilde{\Theta}(n)$ upper and lower bounds due to Nanongkai\cite{Danupon-paths}. Though, the exact version remained open, with the best previous upper bounds remaining at the trivial extreme of $O(m)$, where $m$ denotes the number of the edges. We note that the result of Huang et al. is more general: for $k$ sources, the complexity grows smoothly as $O(n^{3/4} k^{1/2} + n)$; though unfortunately this bound always remains $\Omega(n)$ even for very small values of $k$.  Moreover, it is important to remark that the result of Huang et al.\cite{huang2017distributed} is more general in that it can handle asymmetric distances along an edge---i.e., where the length of the $(s,t)$ edge may differ from that of $(t, s)$---while Elkin's SSSP algorithm assumes the symmetry\cite{elkin2017SSSP}.

\subsection{Our Contribution} 
In this paper, we present algorithms that improve on this state of the art both quantitatively and qualitatively for integer-weighted graphs. We will soon discuss the quantitative aspect, i.e., the improvements in the bounds. The qualitative aspect is the fact that our algorithms extend to networks with asymmetric weights, i.e., when the two directions of an edge can have different weights. Although this asymmetric case is relevant for practical networks, it has gained a far more important motivation recently even when working on graphs with symmetric weights, due to the introduction of the \emph{scaling framework}\cite{huang2017distributed} to distributed computing. The scaling framework massages the weights in a way that simplifies the problem considerably, modulo the potential of making the weights asymmetric. We will discuss this framework in the next section.

We next overview our round complexity improvements. Our basic result gives an SSSP algorithm with complexity $\tO(n^{3/4})$, for graphs with polylogarithmic diameter $D$. This improves on the $\tilde{O}(n^{5/6})$ bound of Elkin\cite{elkin2017SSSP}, and gets closer to the $\Omega(n^{1/2})$ lower bound of Peleg and Rubinovich\cite{Peleg-Rubinovich-1999}, which holds for graphs of diameter $D=O(\log n)$. More generally, the result is as follows:

\begin{theorem}\label{thm:main1} There is a randomized distributed algorithm that computes the exact distances from the source $s$ to all nodes $t \in V$, in $\tilde{O}(n^{3/4}D^{1/4})$ rounds, with high probability.
\end{theorem}

For graphs of large diameter, we present improved versions of this algorithm which achieve complexity $\tO\lp n^{3/4+o(1)}+ \min\{ n^{3/4}D^{1/6},n^{6/7}\}+D\rp $. This compares favorably with Elkin's bound of $\tilde{O}( n^{5/6}+ n^{2/3}D^{1/3} + D )$ in essentially the entire range of parameters. Due to the space limitations, these improvements are deferred to \Cref{section:ImprovedVirtualSSSP}. 
\begin{theorem}\label{thm:main2} There is a randomized distributed algorithm that computes the exact distances from the source $s$ to all nodes $t \in V$, in $\tO\lp n^{3/4+o(1)}+ \min\{ n^{3/4}D^{1/6},n^{6/7}\}+D\rp $ rounds, with high probability.
\end{theorem}

We also show in \Cref{section:MultipleSources} how to extend our algorithm to the case with $\kappa$-sources, providing some improvements on the results of Huang et al.\cite{huang2017distributed} for small $\kappa$ and $D$. 
\begin{theorem}\label{thm:main3} There is a randomized distributed algorithm that computes the exact distances from $\kappa$ given source nodes $s_1$, $s_2$, \dots, $s_\kappa$ to all nodes $t\in V$, in 
	\[\tO \lp \min \left\{  \kappa^{1/2}n^{3/4+o(1)}+ \kappa^{1/3}n^{3/4}D^{1/6} , \kappa^{3/7}n^{6/7} \right\} + D \rp  \]
	rounds, with high probability.
\end{theorem}

\section{Preliminaries}
\label{sec:prelim}



\paragraph{Definitions---Paths, Distances, and Neighborhoods} For a path $P=(v_0,v_1,\ldots,v_h)$ from $v_0$ to $v_h$, we define the \defn{length} of the path $P$ to be the summation of the weights of its edges, i.e., $w(P):=\sum_iw(v_{i-1},v_i)$. Moreover, we say that $P$ has $h$ hops, where $h$ is the number of edges in $P$. A \defn{shortest path} from node $s$ to $t$ is an $s\to t$ path $P$ that minimizes $w(P)$. Its length is denoted $d(s,t)$, also called the \defn{distance} from node $s$ to $t$. Given a source node $s$, the source-wise \defn{$s$-radius} of the graph is equal to $\max_td(s,t)$. Similarly, the \defn{$h$-hop distance} $d^h(s,t)$ is the minimum length of an $s\to t$ path $P$, among all paths that have at most $h$ hops. We define the \defn{$h$-hop distance neighborhood} from a source $s$ as the distances to all vertices $t\in V$ such that $d^h(s,t)=d(s,t)$, i.e., there exists a shortest $s\to t$ path with at most $h$ hops; for all other $t\in V$ with $d^h(s,t)\ne d(s,t)$, their distance is undefined. 

\subsection{Tool 1: The Scaling Framework}
The general case of shortest path problem is hard to tackle directly, while the problem becomes more manageable if the maximum distance from the source $s$ to any other node $t$ is somewhat small. To leverage this, we make use of Gabow's scaling method \cite{gabow1983scaling}. We note that this method was used for the first time in the distributed setting for shortest paths by \cite{huang2017distributed}.  

\paragraph{What do we get from scaling?} The interface of this method is as follows: Given a graph with integer weights in $[0,\Lambda]$, the scaling framework reduces the problem to $\log_2\Lambda$ iterations of computing SSSP on a graph with nonnegative integer weights where in each iteration we have $d(s,t)\le n-1$ for all nodes $t\in V$. That is, in each iteration, the source-wise $s$-radius is bounded by $n-1$. We emphasize that the weights in each of the iterations can be zero, and asymmetric, even if the original weights in the graph were all positive and symmetric. This poses a challenge for standard distributed distance computation methods, which we will discuss in the other tools.  

To summarize, the scaling framework implies that, modulo this challenge of having to handle zero weights and asymmetric weights and an $O(\log\Lambda)$ factor overhead in the running time, it is sufficient to solve SSSP only on graphs satisfying $d(s,t)\le n-1$ for all $t$.

\paragraph{How does scaling work?} Represent each weight $w(u,v)$ in base 2 using $T:=\lf\log_2\Lambda\rf+1$ bits, including leading zeroes if necessary. Let $w_i(u,v)$ denote the integer representing the $i$ most significant bits of $w(u,v)$. Let $b_i(u,v)$ denote the $i$'th most significant bit of $w(u,v)$. The scaling algorithm is as described in \ref{Scaling}.

\mylabel{Scaling}{\texttt{ScalingFramework}}
\begin{algorithm}
\caption{\ref{Scaling}$(s,G=(V,E,w))$}
\small
\textbf{Input}: Every node $v\in V$ knows the weight and direction of each edge $e\in E$ incident to $v$.

\textbf{Output}: Every node $v$ learns $d(v)$, which equals the exact distance $d_w(s,v)$.
\begin{algorithmic}[1]

\State $\tilde d_0(v) \gets 0$ for all nodes $v$

\For {$i=1,\lds,T$ increasing}
  \State Every node $v$ broadcasts $\tilde d_{i-1}(v)$ to its neighbors
  \State Every node $u$ computes $\ell_i(u,v) \gets 2 \tilde d_{i-1}(u) + w_i(u,v) - 2 \tilde d_{i-1}(v)$ for each neighbor $v$
  \State $\de_i(v) \gets$ $\texttt{SSSP}(s,\ell_i)$ \Comment{For each $v\in V$, $\texttt{SSSP}(s,\ell_i)$ informs node $v$ the value of $d_{\el_i}(s,t)$}
  \State $\tilde d_i(v) \gets 2 \tilde d_{i-1}(v) + \de_i(v)$ for all nodes $v$
\EndFor

\State $d\gets \tilde d_{\lc\log_2\Lambda\rc}$

\end{algorithmic}
\end{algorithm}

We refer to \cite{huang2017distributed} for the following lemma, which is itself derived from earlier incarnations of the scaling framework in centralized and parallel settings \cite{gabow1983scaling}.

\BL For each $i\in[T]$, $\ell_i$ and $w_i$ satisfy the following:
(1) For all $(u,v)$, $\ell_i(u,v)\ge0$. (2) For all $t$, $d_{\ell_i}(s,t)\le n-1$. (3) For all $t$, $d_{w_i}(s,t)=2d_{w_{i-1}}(s,t)+d_{\ell_i}(s,t)$. Hence, by induction, $\de_i(v)=d_{\ell_i}(v)$ and $\tilde d_i(v)=d_{w_i}(v)$, and thus $d(v)=\tilde d_{\lc\log_2\Lambda\rc}(v)$ is the correct distance.
\EL

\BC
After $T$ iterations of \texttt{SSSP} on a graph with nonnegative integer weights such that $d(s,t)\le n-1$ for all $t\in V$, \ref{Scaling} correctly outputs all distances.
\EC

\subsection{Tool 2: The ShortRange Algorithm}
Another ingredient in our algorithm is to compute the $h$-hop distances from a given source $s'$ (not necessarily the actual source $s$). For this, we use an algorithm of \cite{huang2017distributed}, called \ref{Short}, which itself is a simple and clever hybrid of the two standard methods, BFS and Bellman-Ford. We next briefly discuss the shortcomings of these methods and then overview \ref{Short}. 

\paragraph{The shortcomings of the standard methods} We can approach the $h$-hop distances problem using the standard methods, namely Bellman-Ford and BFS. However, each of these has a shortcoming, as we discuss next. The Bellman-Ford (B-F) algorithm computes $h$-hop distances in $h$ rounds: in each round, every vertex updates its best known distances and then broadcasts it to each neighbor. However, B-F comes with a downside in edge \textit{congestion}: there may be $\Om(h)$ messages passed along a single edge, which is the largest possible for an algorithm running in $h$ rounds. This is problematic, especially because we will need to compute $h$-hop distances for multiple sources at the same time, one for each node in a chosen subset $V'\subseteq V$. In that case, the high edge congestion of this algorithm becomes a bottleneck. Specifically, if we run B-F from each $s'\in V'$, then an edge may need to pass  $\Om(h|V'|)$ messages, immediately giving a lower bound of $\Om(h|V'|)$ rounds to compute all $h$-hop distances in the \congest model.

On the other hand, if the graph was unweighted, then breadth-first search (BFS) for $h$ rounds would compute $h$-hop distances correctly, while sending at most one message along each edge. Hence, BFS would be a good candidate in terms of edge congestion. However, our graph is weighted and also contains zero-weight edges, which BFS alone cannot solve.

\paragraph{The ShortRange algorithm} To address both of the above issues, \cite{huang2017distributed} introduced the algorithm \ref{Short}, which is a hybrid of BFS and B-F. It computes $h$-hop distances for a source $s'$ in few rounds and with low edge congestion, at a cost of limiting itself only to $t\in V$ satisfying $d^h(s,t)\le\ell$ for some parameter $\ell$. That is, it computes the $h$-hop distance neighborhood restricted to only the nodes $t$ satisfying $d(s,t)=d^h(s,t)\le\ell$. For a node $t$ not satisfying this restriction, the algorithm computes a distance that may overestimate $d(s,t)$, but is never an underestimate. The pseudo-code is deferred to \Cref{app:sec:prelim}, due to space limitations.

\BL
\label{obs:ShortRangeRuntime}
For any parameter $q\ge1$, the ShortRange algorithm runs in $O(\el q+h)$ rounds, and sends $O(h/q)$ messages along each edge. The distances that it computes satisfy the following properties: (1) $\forall t$, $d(t) \ge d_w(s,t)$. (2) If the shortest $s\to t$ path has at most $h$ hops and length at most $\el$, then $d(t)=d_w(s,t)$.
\EL


%

\subsection{Tool 3: Virtual Node Sampling}

Following the steps of \cite{ullman1991high} and previous distributed shortest path algorithms \cite{Lenzen-PattShamir,Danupon-paths, elkin2017SSSP, huang2017distributed}, our algorithm samples a set of \defn{virtual} nodes $V'\s V$ as follows: every node independently joins the set with probability $k/n$, for a parameter $k$. We also add $s$ to $V'$, i.e., $s$ joins $V'$ with probability $1$ instead. A simple Chernoff bound shows that the size of $V'$ is $O(k+\log n)$ w.h.p. Furthermore, $V'$ satisfies the following crucial property:

\BL \label{lemma:Virtual}
W.h.p., for every $t\in V$, there is a shortest $s\to t$ path that has a virtual node inside every consecutive $\Th(n\log n/k)$ vertices on the path. We call this a good $s\to t$ shortest path. In particular, the last virtual node on this path is within $O(n\logn/k)$ hops from node $t$.
\EL

\subsection{Tool 4: Distributed Scheduling of Algorithms}\label{section:Scheduling}
\vspace{-0.1cm}
\newcommand{\dilation}{$\mathsf{dilation}$\xspace}
\newcommand{\congestion}{$\mathsf{congestion}$\xspace}
At certain points, our main algorithm will be calling a set of distributed subprocedures \textit{in parallel}. That is, the computations of any two distinct subprocedures do not depend on each other. We cannot actually run all of these subprocedures in parallel, since many subprocedures might need to send a message through the same edge $e\in E$ simultaneously, but the \congest model only allows an $O(\log n)$-bit sized message per round through $e$. On the other hand, we do not want to run these subprocedures sequentially, one after another. Our solution is to \textit{schedule} the distributed algorithms, following \cite{Scheduling}, based on two parameters called \dilation and \congestion.

Formally, consider a set of distributed algorithms $A_1,\ldots,A_k$ that run independently of one another. We define \dilation as the maximum running time of any algorithm $A_i$, and \congestion as the maximum number of times a single edge is used over all $k$ algorithms. Formally, define $b_{e,i}$ as the number of times the algorithm $A_i$ sends a message along edge $e\in E$, and let $\mathsf{congestion} := \max_e \sum_{i=1}^kb_{e,i}$. Note that \dilation and \congestion are both lower bounds on the time it takes to run all the algorithms. It turns out that the upper bound nearly matches these lower bounds.

\BT\cite{Scheduling}\label{thm:Scheduling}
Given a set of independent algorithms with parameters \dilation and \congestion, we can run them all in time $\tO(\textsf{dilation}+\textsf{congestion})$, w.h.p.
\ET

\subsection{Tool 5: Additive Approximation of SSSP} \label{section:AdditiveApprox}
\vspace{-0.1cm}
In our exact SSSP computation, we make use of an \textit{approximate} SSSP $\tilde d$ that is accurate up to some \textit{additive error} $\al$; that is, for all nodes $t$, we have $d(s,t)\le \tilde d(s,t)\le d(s,t)+\al$. To our knowledge, this is the first time a distributed exact shortest path algorithm relies on an approximate algorithm as a precomputation step.
To compute this $\al$-additive approximate SSSP, we use a method very similar to (a part of) the $(1+\eps)$-multiplicative approximation algorithm of Nanongkai \cite{Danupon-paths}. It is notable that we do not utilize the hop-set idea of Nanongkai's algorithm, since hop-sets do not translate to directed graphs. Since Nanongkai \cite{Danupon-paths} does not explicitly state the dependencies on the error term $\e$, we present, for completeness, the full algorithm and its analysis in the Appendix. Finally, since the scaling framework guarantees us that the $s$-radius is at most $n-1$, we can take $\e:=\al/n$ in the $(1+\e)$-multiplicative approximation to obtain an $\al$-additive approximation.

\begin{restatable}{lemma}{ApproximateSSSP}\label{lemma:AdditiveErrorSSSP}
Suppose that the network graph has $s$-radius at most $n-1$. Then, we can compute $\al$-additive approximate SSSP in $\tO(\f{n^2}{\al k}+kD)$ time, and with $\tO(k)$ messages sent along each edge.
\end{restatable}
\vspace{-0.1cm}
\section{Main Algorithm}
\label{sec:main}
\vspace{-0.2cm}
\subsection{Algorithm Overview}
In this section, we present our algorithm that proves \Cref{thm:main1} and computes distances from source $s$ to all nodes $t \in V$, in $\tilde{O}(n^{3/4}D^{1/4})$ rounds. We start with a high-level description of the algorithm. On the outer level, this algorithm runs the \ref{Scaling} for $O(\log\Lambda)$ iterations, solving an SSSP problem in each iteration. Recall that, on every iteration, \ref{Scaling} reweights the edges on the graph so that the $s$-radius, defined to be $\max_td(s,t)$, is at most $n-1$. Once SSSP is solved on this reweighted graph, \ref{Scaling} proceeds to the next iteration, reweighting the edges again to have $s$-radius at most $n-1$. This happens for $O(\log\Lambda)$ iterations, after which each node can locally deduce its correct distance from node $s$. Fix an iteration of \ref{Scaling}, and let $G=(V,E)$ denote the reweighted graph with $s$-radius at most $n-1$. We next describe our algorithm \ref{Small}, which solves SSSP on this graph.



\paragraph{Virtual Graph}
Following \Cref{lemma:Virtual}, every node becomes a virtual node with probability $k/n$, for a paremeter $k\leq n$ to be fixed, except for $s$ which becomes virtual with probability $1$.
\medskip

To build intuition for the rest of the algorithm, let us first consider the task of computing $d(s,u')$ for only the virtual nodes $u'\in V'$. Let $h:=\Theta(n\log n/k)$, and consider the \textit{virtual graph} $G'=(V',E')$ with vertex set $V'$ and an edge between each $u',v'\in V'$ of weight $d(u',v')$ if the shortest $u'\to v'$ path has $\le h$ hops. The graph $G'$ is called virtual because the edges in $E'$ are not necessarily edges in the network, making direct communication between edges in $E'$ impossible.  Observe that, by \Cref{lemma:Virtual}, running SSSP from $s\in V'$ in the virtual graph $G'$ will, w.h.p., result in the same distances to each virtual node as running SSSP from node $s$ in $G$. Therefore, a natural next step for an algorithm is to compute all edges in $E'$; specifically, for every edge $(u',v')$, we want  node $v'$ to know its weight and direction.\footnote{Since communication on the network is bidirectional, it turns out that we can also allow node $u'$ to learn edge $(u',v')$. However, this will not be necessary.}

However, our algorithm will not compute all edges in $E'$, but only a subset of the edges. Define an edge $(u',v')\in E'$ to be \defn{light} if $d(u',v') \le \el$ for a parameter $\el$, and \defn{heavy} otherwise. Let $G'_L=(V',E'_L)$ and $G'_H=(V',E'_H)$ be $G'$ with only light edges and heavy edges in $E'$, respectively. The algorithm will compute all light edges in $E'_L$, so that for each directed edge $(u',v')\in E'_L$, the endpoint $v'$ knows that edge. For the heavy edges in $E'_H$, computing them can be time-consuming; thus, the algorithm will take another approach. We next discuss these two cases separately.

\paragraph{Light Edges}
To compute all light edges, we run \ref{Short}$(u',G,h,\el,q)$ for each virtual node $u'\in V'$, for some parameter $q$. For each pair $u',v'\in V'$, let $d_{u'}(v')$ be the value of $d(v')$ returned by \ref{Short}$(u',G,h,\el,q)$. If $d_{u'}(v')<\infty$, then $v'$ can now learn a virtual edge $(u',v')$ with weight $d_{u'}(v')$. For each light edge $(u',v')$ in $E'_L$, $v'$ learns the virtual edge $(u',v')$ with correct weight $d(u',v')$, due to the guarantee of \ref{Short}. There may be other extraneous virtual edges learned by virtual nodes that overestimate their distance, but, as we will see, they will not affect the output.

\paragraph{Heavy Edges}
Note that the only important virtual edges in $E'_H$ are those that belong to  the shortest path tree from node $s$ in $G'$. Hence, it suffices to limit our attention to the heavy edges $(u',t')\in E'_H$ that belong to a shortest $s\to t'$ path in $G'$, for some node $t'$. 
 In this case, we leverage the fact that $d(s,t')>d(s,u')+\ell$. We partition $V$ into $\lf n/\el\rf+1$ \textit{buckets} such that bucket $i\in[0,\lf n/\el\rf]$, denoted $V_i$, has all  nodes $v$ with $d(s,v)\in[i\el,(i+1)\el)$. We will discuss later how to compute the buckets.\footnote{At first glance, this might seem troublesome, since it requires knowing the distances beforehand.} Observe that the bucket containing $u'$ has a smaller index than the one containing $t'$. The algorithm will compute the correct distances to each bucket in increasing bucket index $i$, with the invariant that before processing bucket $V_i$, all  nodes in $V_0$ through $V_{i-1}$ know their correct distance from node $s$. To process bucket $V_i$, the algorithm first executes a single B-F to depth $h$,\footnote{In the algorithm pseudocode, we run an algorithm \ref{Extend} instead. The reader is encouraged to replace the \ref{Extend} calls with calls of B-F to depth $h$; see Remark~\ref{remark:Extend}. We use B-F for this discussion to simplify this overview.} then runs a \emph{virtual SSSP}, and then runs another B-F to depth $h$. We will soon explain the \emph{virtual SSSP} algorithm and the difficulties in it, as it needs to be run on a virtual graph. By a B-F to depth $h$, we mean initializing the B-F with distances $d(v)$ for all $v\in V_i$, and then running B-F, which broadcasts and updates distances for $h$ rounds. Going back to our consideration of a heavy edge $(u',t')\in E'_H$, note that, right before the algorithm processes the bucket containing $t'$, the bucket containing $u'$ has already been processed, and $u'$ knows its correct distance. Then, when the algorithm processes the bucket containing $t'$, in the first B-F to depth $h$, node $u'$ will propagate its distance to node $t'$, informing $t'$ of its correct distance from $s$.

\paragraph{Virtual SSSP}
We now discuss how we run the virtual SSSP when processing each bucket. Suppose the algorithm is currently processing bucket $V_i$. By the invariant, all nodes in $V_1$ through $V_{i-1}$ already know their correct distance, and the objective is now for those of bucket $V_i$ to learn their distance. Note that a B-F to depth $h$ alone is not enough. This is because edges can have weight zero and thus, a virtual node in $V_{i}$ can be a large number of hops away from any node in $V_1$ through $V_{i-1}$, even if the total weighted distance is small. 

Consider some virtual node $t'\in V_{i}$ and the good $s\to t'$ path $P_{t'}$, guaranteed to exist by \Cref{lemma:Virtual}. Although there may be many virtual nodes on $P_{t'}$ belonging to bucket $V_{i}$, the key observation is that, of the virtual nodes in  $P_{t'} \cap V_{i}$ ordered by $P_{t'}$, the \textit{first} node  knows its correct distance, and for the remaining nodes, there is an edge in $E'_L$ from each one to the next. For this \textit{first} virtual node $v'\in P_{t'}\cap V_i$, we will show that, after the first B-F in processing bucket $V_i$---as mentioned above when discussing heavy edges---node $v'$ will learn its correct distance.  Afterward, we would like to propagate the distance of $v'$ along the edges in $E'_L$ by solving a SSSP problem on $G'_L$. We define this SSSP instance as follows: we begin with the subgraph $G'_L$ induced by $\{s\}\cup V_i$, and for each node $v'\in V_i$ that has computed some (possibly incorrect)\ distance $d(v')$, we add the edge $(s,v')$ with weight $d(v')$. We can think of each additional edge $(s,v')$ as \textit{shortcutting} some path from $s$ to $v'$.   We now run SSSP on this subgraph, but we cannot communicate along the virtual edges in $G'_L$ directly, since they may not exist in the network. Hence, we instead run a \textit{virtual SSSP} on the subgraph of $G'_L$ induced by $V'\cap V_i$. This \ref{Virtual} is described in \Cref{section:VirtualSSSP}. Together, \ref{Virtual} and the B-Fs correctly compute all distances from node $s$ to the virtual nodes.

\paragraph{Computing the Buckets} We now discuss how to compute the buckets $V_i$. We will actually compute a set of nodes $\tilde V_i$ that contains $V_i$, so that the subgraph of $G'_L$ induced by $V'\cap\tilde V_i$ contains all vertices and edges in the target subgraph. However, we also need that the $\tilde V_i$ are, on average, not much larger than the $V_i$, in order to maintain a fast algorithm.

First, we use the additive approximate distance algorithm of \Cref{section:AdditiveApprox} to compute $\el$-additive approximate distances from node $s$ to every other node. Since the buckets have distance intervals of size $\el$, each node $v'\in V'$ can pinpoint down two buckets, one of which contains $v'$. Then, define $\tilde V_i$ to include every $v'\in V'$ that guesses $V_i$ as one of its two possible buckets. The total size of all $V'\cap \tilde V_i$ is at most twice the total size of all $V'\cap V_i$, so the asymptotic running time is unaffected.

\paragraph{Extending the Distance Computation to Non-Virtual Nodes} By \Cref{lemma:Virtual}, every node $t\in V$ has a shortest $s\to t$ path that contains a virtual node $v'\in V'$ within its last $h$ vertices, w.h.p. Consider the iteration when the algorithm processes the bucket containing $v'$. Node $v'$ learns its correct distance $d_w(s,v')$ right after the virtual SSSP computation.  Then, in the second B-F to depth $h$, node $v'$ propagates its distance to node $t$, and node $t$ learns its correct distance. Thus, node $t$ learns its correct distance after the algorithm processes the bucket containing $v'$, which either equals or appears earlier than the bucket containing $t$. Hence, once the algorithm finishes bucket $V_i$, all nodes in buckets $V_1$ through $V_i$ learn their correct distance, satisfying the invariant. 

\subsection{The Algorithm}

In this section, we give a detailed description of our main algorithm, which computes exact shortest paths in $\tO(n^{3/4}D^{1/4})$ time, proving \Cref{thm:main1}. The main algorithm, called \ref{Main}, first computes an $\el$-additive approximate SSSP from source $s$; recall that this approximate computation allows each node to guess which bucket it belongs to. Then, it runs \ref{Scaling}, reducing the problem to $O(\log\Lambda)$ computations of SSSP, each time on a graph with $s$-radius at most $n-1$. On each iteration, we call the subprocedure \ref{Small}, which is the SSSP algorithm with the promised $s$-radius bound. We will defer the pseudo-code of \ref{Main} to \Cref{app:sec:main}.

\paragraph{Parameters} The algorithm will maintain the parameters $k,h,\el,q$, as described below. For an initial read, we encourage the reader to assume that $D=\tO(1)$ and ignore the $D$ factors.

\BI
\im $k$ is the desired number of virtual nodes. Every node becomes virtual with probability $k/n$, except for $s$ which is virtual. W.h.p., we sample $O(k+\log n)$ virtual nodes.
\im $h$ is a parameter given to the \ref{Short} algorithm, a bound on the number of hops. We set $h=\Th(n\log n/k)$ so that, by \Cref{lemma:Virtual}, w.h.p.\ for every node $t$, there is an $s\to t$ path with a virtual node inside every consecutive $\Th(n\log n/k)$ vertices on the path.
\im $\el$ represents four quantities, which we will observe are closely related to one another. First, it is the parameter into the \ref{Short} algorithm, a bound on the maximum distance. Second, it is the parameter in the initial approximate shortest paths computation; that is, the algorithm initially computes $\el$-additive approximate shortest path distances from source $s$. Third, it is the threshold such that an edge in the virtual graph is called light if its weight is at most $\el$, and heavy otherwise. Lastly, it is the width of each bucket: bucket $i$ represents all nodes whose distance from $s$ is within $[i\el,(i+1)\el)$.
\EI

These parameters are summarized below, along with their values in the $\tO(n^{3/4}D^{1/4})$ algorithm.
\begin{center}
\small
    \begin{tabular}{| l | p{2.5cm} | p{12cm} | }
    \hline
    Param. & Value & Description \\ \hline
                $k$ & $n^{3/4}D^{-3/4}$ & The number of virtual nodes we want to sample.  \\ \hline
    $h$ & $\Th(n\log n/k)=\tO(n^{1/4}D^{3/4})$ & The parameter for the number of hops in the \ref{Short} algorithm. \\ \hline
                $\el$ & $n^{1/2}D^{1/2}$ & The parameter for the maximum distance in the \ref{Short} algorithm. Also, the error of the initial additive approximation. \\ \hline
        $q$ & $n^{1/4}D^{-1/4}$ & The optimization parameter for the \ref{Short} algorithm. \\ \hline
    \end{tabular}
\end{center}

\paragraph{Algorithm Description} Below, we present the pseudocodes for \ref{Small} and \ref{Extend}.


\mylabel{Small}{\texttt{SmallWeightSSSP}}
\begin{algorithm}[H]
\small
\caption{$\ref{Small}(s,G=(V,E,w),\tilde d,k,\el,q)$}

\noindent\textbf{Input}: Every node $v$ knows the weight and direction of each edge $e\in E$ incident to $v$ and the $\el$-additive distance estimate $\tilde d(v)$. We have the promise that $d_w(s,t)\le n-1, \forall t\in V$. 

\noindent\textbf{Output}: Every node $v$ learns $d(v)$, which equals the exact distance $d_w(s,v)$.
\begin{algorithmic}[1]


\State\label{line:ShortRangeBegin}Every node becomes virtual with probability $k/n$, except for $s$, which becomes virtual with probability $1$. Let $V'\s V$ be the set of virtual nodes. {\tt\color{gray}\color{gray} // We have $s\in V'$ and, w.h.p., we sample $O(k+\log n)$ virtual nodes.}

\State\label{line:ShortRangeEnd}Set $h\gets \Th(n\log n/k)$. For each $s'\in V'$, let $d_{s'}$ be the result of $\ref{Short}(s',G,h,\el,q)$. Each $v\in V$ knows $d_{s'}(v)$ for each $s'\in V'$.

\State\label{line:BuildGraph}Build graph $G'=(V',E',w')$ with directed edge $(u',v')\in E'$ having weight $d_{u'}(v')$, for each pair $(u',v')\in V'\times V'$ satisfying $d_{u'}(v')<\infty$. Every $v'\in V'$ knows each directed edge $(u',v')\in E'$ and its weight.

\State
  Initialize distances $d(s)\gets0$, $d(v)\gets\infty$ for all $v\in V\setminus s$.


\For {$i=0,\ldots,\lf n/\el\rf$} {\tt\color{gray} // Iterate over buckets $V_0,\ldots,V_{\lf n/\el\rf}$.}
  \State\label{line:FirstExtend}
    Run $\ref{Extend}(G,d,h,\el,i)$, and let $d$ be the updated distance that each node knows. {\tt\color{gray} // Equivalently, run a B-F to depth $h$.}
  \State
    Define $\tilde V'_i$ as $\{ v'\in V' : \tilde d(v')\in[i\el,(i+2)\el) \}$. {\tt\color{gray} // $\tilde V_i'$ contains all virtual nodes that guess bucket $i$ as one of their two guesses. In particular, $V'\cap V_i \s \tilde V_i'$.}
  \State
    Consider the subgraph of $G'$ induced by $\{s\}\cup \tilde V'_i$. For each $v'\in\tilde V_i'$ with $d(v')<\infty$, add an edge $(s,v')$ of weight $d(v')$ to this subgraph. Call the resulting graph $\tilde G_i'$. {\tt\color{gray}// Note that every node $v'\in\tilde V_i'$ still knows every edge $(u',v')$ in $\tilde G_i'$.}  \State \label{line:VirtualSSSP}
    Run $\ref{Virtual}(s,\tilde G_i')$, and let $\tilde d_i'$ be the resulting distance that each node learns.
  \State
    Every $v'\in V'$ updates $d(v')\gets \min\{d(v'),\tilde d_i'(v')\}$.
  \State\label{line:SecondExtend}
    Run $\ref{Extend}(G,d,h,\el,i)$, and let $d$ be the updated distance that each node knows. {\tt\color{gray} // Equivalently, run a B-F to depth $h$.}

\EndFor

\end{algorithmic}
\end{algorithm}



\mylabel{Extend}{\texttt{Extend}}
\begin{algorithm}[H]
\small
\caption{$\ref{Extend}(G,d,h,\el,i)$}
\noindent\textbf{Input}: Every node $v$ knows its current distance estimate $d(v)$.

\noindent\textbf{Output}: Every node $v$ updates its distance estimate $d(v)$.
\begin{algorithmic}[1]

  \For{$h$ rounds} \label{line:RunBF} {\tt\color{gray} // Run a \textit{restricted} B-F to depth $h$. See Remark~\ref{remark:Extend}.}
    \State
      Every node $u$ sends $d(u)$ to all neighbors $v$ with $d(u)+w(u,v)\in[i\el,(i+1)\el)$
    \State
      Every node $v$ sets $d(v)=\min\{d(v),d(u)+w(u,v)\}$ for each $d(u)$ received from neighbor $u$
  \EndFor
\end{algorithmic}
\end{algorithm}

\begin{remark}\label{remark:Extend}
        On line~\ref{line:FirstExtend} and line~\ref{line:SecondExtend}, we can replace the algorithm \ref{Extend} with a B-F to depth $h$ for the same correctness and running time of \ref{Small}.\footnote{The algorithm \ref{Extend} is actually a \textit{restricted} B-F to depth $h$, in that only nodes that satisfy a certain property can broadcast their distance.} The reader is encouraged to do so for a simpler algorithm and the same proof of correctness.
        However, we need \ref{Extend} for the multiple sources case in \Cref{section:MultipleSources}.
\end{remark}

\subsection{Analysis}
First, we prove the aforementioned invariant: after the algorithm processes bucket $V_i$, all nodes in buckets $V_1$ through $V_i$ know their correct distance.

\BL\label{lemma:VirtualCorrectDistances}
At the end of each iteration $i$ of the \textbf{for} loop, $d(t)=d_w(s,t)$ for all $t$ such that $d(t)\in[0,(i+1)\el)$. 
Hence, at the end of \ref{Small}, $d(t)=d_w(s,t)$ for all $t\in V$. In other words, every node knows its correct distance from node $s$.
\EL
\BP
First, consider the base case iteration $0$ and a node $t$ with $d(s,t)\in[0,(i+1)\el) = [0,\el)$. Consider the shortest $s\to t$ path $P_{t'}$ satisfying the properties in \Cref{lemma:Virtual}, and let $t'$ be the last virtual node on $P_t$, possibly $s$ itself. Since each virtual node $v'$ on the path satisfies $d(s,v')\in[0,\el)$, its approximate distance satisfies $\tilde d(s,v')\in[0,2\el)$, so $v'\in\tilde V_0'$. Also, w.h.p., consecutive virtual nodes in $P_{t}$ are within $h=O(n\log n/k)$ hops away from each other. Since they are also distance $\le \el$ apart, they are connected by edges in $E'$, and $\ref{Virtual}(s,\tilde G_i')$ (line~\ref{line:VirtualSSSP}) correctly computes $\tilde d_i'(t')=d(t')$. Here, we use the fact that edges in $\tilde G_i'$ can only be overestimates of their true distances in $G$, so that we can never have $\tilde d_i'(t')< d(t')$. Finally, since $t'$ and $t$ are $\le h$ hops apart on $P_t$, and since every node in between $t'$ and $t$ has distance from node $s$ in the range $[0,\el)$, the second $\ref{Extend}$ algorithm (line~\ref{line:SecondExtend}) correctly computes $d(v)$ for each $v$ in between $t'$ and $t$ on $P_t$. In particular, $t$ will know its correct distance $d(t)$.

We now apply induction on $i$. If node $t$ satisfies $d(t)\in[0,i\el)$, then the statement $d(t)=d_w(s,t)$ has already been proved on iteration $i-1$, so assume that $d(t)\in[i\el,(i+1)\el)$. Take the shortest $s\to t$ path $P_{t}$, let $u$ be the last node on $P_t$ satisfying $d(s,u)<i\el$, let $v$ be the next node after $u$ on $P_{t}$, and let $v'$ be the next \textit{virtual} node after $u$ on $P_{t}$. Let $t'$ be the last virtual node on $P_t$, possibly $v'$ or $s$ itself. Suppose that $d(u)\in[j\el,(j+1)\el)$ for some $j<i$. By induction, $d(u)$ was correctly computed at the end of iteration $j$. Observe that all nodes on $P_{t}$ from $v$ to $v'$ have their distance from node $s$ in the range $[i\el,(i+1)\el)$, and moreover, $v$ and $v'$ are $\le h-1$ hops apart on $P_{t}$ w.h.p. Therefore, for the $h$ iterations of the first $\ref{Extend}$ (line~\ref{line:FirstExtend}) on iteration $i$, $u$ will first send $d(u)+w(u,v)$ to $v$, and then $h-1$ rounds later, $v'$ will update $d(v')$ to be its correct distance.
 The rest of the argument follows similarly to the base case. All virtual nodes in $P_{t'}$ from $v'$ onwards belong to $\tilde V'_i$, and since they are $\le h=O(n\log n/k)$ hops away from each other, they are connected by edges in $\tilde G_i'$. Therefore, the path consisting of $s,v'$, and all virtual nodes after $v'$ on $P_{t'}$ has length exactly $d(t')$ in $\tilde G_i'$, and the \ref{Virtual} (line~\ref{line:VirtualSSSP}) of iteration $i$ computes $d'(t')$ correctly. Finally, using the same argument as in the base case,  the second $\ref{Extend}$ (line~\ref{line:SecondExtend}) correctly computes $d(t)$.
\EP

\BL\label{lemma:SSSPRunningTime}
\ref{Small} runs in $\tO\lp \el q+\f nq+kD+\f{n^2}{\el k}\rp = \tO(n^{3/4}D^{1/4})$ time.
\EL
\BP
Computing an $\el$-additive approximate SSSP takes $\tO(n^2/(\el k) + kD)$ time by \Cref{lemma:AdditiveErrorSSSP}. We use the same value of $k$ for $\el\texttt{-AdditiveErrorSSSP}$ as we do for \ref{Small}.

We run \ref{Short}$(\cd,\cd,h,\el,q)$ for $k$ nodes, with $h=O(n\log n/k)$, and each node congests a given edge $O(h/q)$ times, so the total congestion is $O(k(n\log n/k)/q)=O(n\log n/q)$. The dilation is $O(\el q + h) = O(\el q + n\log n/k)$. We can schedule all the \ref{Short} algorithms in time $\tO(\textsf{dilation}+\textsf{congestion})$ by \Cref{thm:Scheduling}, giving time $\tO(\el q+n/k+n/q)$.

For $O(n/\el)$ iterations, we run \ref{Virtual} on a graph with $|\tilde V'_i|+1$ vertices, which takes $O((|\tilde V'_i|+1)D\log n)$ time by \Cref{lemma:VirtualSSSP}, and $\ref{Extend}(\cd,\cd,h,\cd,\cd)$, which takes time $h+1=O(n\log n/k)$. Every $v'\in V'$ belongs to at most $2$ sets $\tilde V_i'$, so $\sum_i|\tilde V_i'|\le2k$ and the total running time for \ref{Virtual} and B-F is $\tO(kD + n/\el\cdot n/k)$.

Setting $k:=n^{3/4}D^{-3/4}$, $\el:=n^{1/2}D^{1/2}$, $q:=n^{1/4}D^{-1/4}$ gives the desired time.
\EP

\subsection{SSSP on Virtual Graph}\label{section:VirtualSSSP}
From \ref{Small}, we have the following setting: there is a virtual graph $G'=(V',E',w')$ such that $d_{w'}(s,v')\le n-1$ for all $v'\in V'$, and for each edge $(u',v')\in E'$ directed from $u'$ to $v'$, only $v'$ knows its weight and direction. Recall that since edges are virtual, they may not exist in the communication network, so we cannot pass information directly along edges in $E'$. For \ref{Virtual}, our goal is to inform each virtual node $v'\in V'$ of its exact distance $d_{w'}(s,v')$.


\mylabel{Virtual}{\texttt{VirtualSSSP}}
\paragraph{Algorithm \ref{Virtual}}
Let $k:=|V'|$. The algorithm \ref{Virtual} first runs \ref{Scaling} on $G'$, reducing the problem to $\log n$ iterations of SSSP with the guarantee $d(s,v')\le k-1$. After each iteration, instead of local broadcasts from each node, we let every node $v'\in V'$  broadcast its computed $d(v')$ to all other nodes, which takes $O(k+D)$ time using standard pipelining techniques on a BFS tree, so that weights on the next iteration can be computed.

The algorithm for each iteration of \ref{Scaling} is promised that $d_{w'}(s,v')\le k-1,  \forall v'\in V'$. We now discuss this algorithm, which we name \ref{SmallVirtual}; its pseudocode is deferred to \Cref{app:sec:main}. It is instructive to first consider this algorithm in a parallel computation setting, as follows. For each $i\in[0,k-1]$, initialize a queue $Q_i$ to be used in the algorithm. To begin, insert the value $(s,0)$ to queue $Q_0$.  We process the queues in increasing order $Q_0,Q_1,\ldots,Q_{k-1}$, continuing onto the next queue whenever the current queue is empty. When processing $Q_i$, we (1) mark each $u'\in Q_i$ as completed, (2) write $d(s,u')$ for each $u'\in Q_i$ in (centralized) memory, and (3) for each unmarked $v'$ with an edge $(u',v')\in E'$, remove $v'$ from its current queue (if any) and insert it to $Q_j$ with $j:=i + \min\limits_{(u',v')\in E'}w(u',v')$. Note that we may have to process $Q_i$ again, since there may be zero-weight edges.
It is straightforward to argue that this algorithm is correct. Furthermore, since at least one node is marked every iteration, the algorithm runs in $O(k)$ parallel rounds.

Finally, we translate this algorithm to a distributed setting. Instead of writing each completed $u'\in Q_i$ to memory, we broadcast $d(s,u')$ to all other nodes in time $O(|Q_i|+D)$ so that all unmarked nodes can locally compute $\min\limits_{(u',v')\in E'}w(u',v')$. To figure out when to advance to the next queue, we continue to process the current queue $Q_i$ until there is a round with no broadcasts. At that point, all nodes can advance to queue $Q_{i+1}$ in a synchronized manner.

\BL\label{lemma:VirtualSSSP}
\ref{SmallVirtual} on a virtual graph $G'=(V',E')$ takes $O(|V'|D)$ rounds, and sends $O(|V'|)$ along each edge. Hence, \ref{Virtual} on a virtual graph $G'=(V',E')$ takes $O(|V'|D\logn)$ rounds, and sends $O(|V'|\logn)$ messages along each edge.
\EL

\BP
Every node broadcasts once, so the total number of broadcasts is $O(|V'|)$. The while loop terminates in $O(|V'|)$ rounds, since in each round, either some node broadcasts (and then never again), or $i$ increases. Every broadcast can be executed in $O(D)$ time through an auxiliary BFS tree, so in total, everything takes $O(|V'|D)$ time. Similarly, every broadcast congests each edge of the BFS tree once, so the total congestion is $O(k)$.
\EP


\paragraph{Acknowledgement} 
The first author is grateful to Michael Elkin for explaining his result\cite{elkin2017SSSP} and discussions about potential improvement directions, and to Danupon Nanongkai for sharing an earlier manuscript of \cite{huang2017distributed}.

\bibliographystyle{alpha}
\bibliography{ref}
\appendix

\section{Missing Details of \Cref{sec:prelim}}
\label{app:sec:prelim}

\mylabel{Short}{\texttt{ShortRange}}
\begin{algorithm}
\small
\caption{$\texttt{ShortRange}(s,G=(V,E,w),h,\el,q)$}
\textbf{Input}: Source $s$, network graph $G=(V,E,w)$, and parameters $h,\el,q$. Every node $v$ knows the weight and direction of each edge $e\in E$ incident to $v$.

\textbf{Output}: Every node $v$ learns $d(v)$, which satisfies $d(v)\ge d(s,v)$. If $v$ is within the $h$-hop distance neighborhood and satisfies $d(s,v)\le\el$, then $d(v)=d(s,v)$.
\begin{algorithmic}[1]

\State $d(s) \gets 0$, $d(t)\gets\infty$ for all $t\ne s$
\State $w'(u,v) \gets \max\{w(u,v), 1/q\}$ for each $(u,v)$ \Comment{$w'(u,v)=1/q$ if $w(u,v)=0$ and $w(u,v)$ otherwise}
\For {$i=0,\ldots,\el q+h$}\label{line:BFSStart} \Comment{BFS}
  \State Every node $u$ with $d(u)=i/q$ broadcasts $d(u)$ to its neighbors
  \State\label{line:BFSEnd}If node $v$ receives $d(u)$, updates $d(v) \gets \min\{d(v), d(u)+w'(u,v)\}$
\EndFor

\State\label{line:AfterBFS}$d(v) \gets \lf d(v) \rf$ for all $v$

\For {$h$ iterations}\label{line:BFStart} \Comment{Bellman-Ford (B-F)}
  \State Every node $u$ broadcasts $d(u)$ \textbf{unless} (1) $d(u)$ has been broadcast previously, or (2) $u$ has broadcast $\lf h/q\rf$ times already
  \State\label{line:BFEnd}If node $v$ receives $d(u)$, updates $d(v) \gets \min\{d(v), d(u)+w(u,v)\}$
\EndFor

\State \Return $d$

\end{algorithmic}
\end{algorithm}

\begin{proof}[Proof of \Cref{obs:ShortRangeRuntime}]
First, we show that $d(t) \ge d_w(s,t)$ for all $t$. Let $d'(v)$ be the value of $d(v)$ right after line~\ref{line:AfterBFS} of \ref{Short}. Since the weights $w'$ are overestimates of $w'$, we have $d'(t)\ge d_w(s,t)$. By the properties of B-F, if a node $t$ has value $d(t)$ at the end of the algorithm, then there is a path $P$ from some node $v\in V$ to $t$ such that $d(t) = d'(v) + w(P)$. Therefore, $d(t) = d'(v)+w(P) \ge d_w(s,v)+w(P) \ge d_w(s,t)$.

Now suppose that the shortest $s\to t$ path $P$ has $\le h$ hops and length $\le \el$. We have $d_{w'}(s,t)\le d_w(s,t)+h/q\le \el+h/q$, since this path has $\le h$ edges of length $0$ which add $\le h/q$ additional length according to $w'$. Moreover, every node $v\in P$ satisfies $d_{w'}(s,v)\le d_w(s,v)+h/q\le \el+h/q$.
The BFS (lines~\ref{line:BFSStart}--\ref{line:BFSEnd}) computes $d(v)=d_{w'}(s,v)$ for all $v$ satisfying $d_{w'}(s,v)\le (\el q+h)/q=\el+h/q$, so for all $v\in P$, $d(v)=d_{w'}(s,v)$ by line~\ref{line:BFSEnd}.
Observe that if the B-F (lines~\ref{line:BFStart}--\ref{line:BFEnd}) did not have restriction (2), then all distances within $h$ hops would be correctly computed. However, after line~\ref{line:AfterBFS}, every node $v\in P$ has $d(v)=\lf d_{w'}(s,v)\rf\le d_w(s,v)+\lf h/q\rf$, and since every broadcast results from decreasing the estimated $d(v)$ by $\ge1$, each node $v$ will indeed broadcast its correct $d(s,v)$ within $\lf h/q\rf$ broadcasts. In particular, $d(t)$ will be the correct $d_w(s,t)$ at the end of the algorithm.
\end{proof}

\begin{proof}[Proof of \Cref{lemma:Virtual}]
For each node $t\in V$, fix a shortest $s\to t$ path $P_t=(s=v_0,\ldots,v_h=t)$, and let $C:=cn\ln n/k$ for a constant $c$. If $h<C$, then the first part of the statement is vacuous, and the second part is immediate since $s$ is a virtual node within $C$ hops from node $t$. 

If $h\ge C$, then for each $i\in[h-C+1]$, the probability that none of the nodes $v_i,\ldots,v_{i+C-1}$ is virtual is exactly \[(1-k/n)^C=(1-k/n)^{cn\ln n/k} < \exp(k/n \cd cn\ln n/k)=n^{-c}.\] Taking a union bound over all values of $i$ and all $n-1$ paths $P_t$ gives the w.h.p.\ result.
\end{proof}

\paragraph{Approximate SSSP, proving \Cref{lemma:AdditiveErrorSSSP}}

Next, we prove \Cref{lemma:AdditiveErrorSSSP}, restated below.

\ApproximateSSSP*

To compute an $\al$-additive approximation on a graph with $s$-radius $\le n-1$, it suffices to compute a $(1+\al/n)$-multiplicative approximation, i.e., $d(s,t)\le\tilde d(s,t)\le(1+\al/n)d(s,t)$. Let $\e:=\al/n$, so that we are looking for a $(1+\e)$-multiplicative approximation (with running time dependent on $\e$).

First, we use the algorithm of \cite{Danupon-paths} to compute $h$-hop distances.
The following is a restatement of Theorem 3.2 from \cite{Danupon-paths} with the added dependence on $\e$ in the running time.
\BT
\ref{BoundedHopSSSP} computes $(1+\e)$-multiplicative approximate $h$-hop distances in $\tilde O(\e^{-1}h)$ time, and $\tilde O(1)$ messages are sent along each edge.
\ET
\BP
For proof of correctness, we refer the reader to Theorem 3.2 of \cite{Danupon-paths}. Note that since $G$ has $s$-radius $\le n-1$, the range $[\lc\log(n-1)\rc]$ suffices in line~\ref{line:ForAllI} of \ref{BoundedHopSSSP}, as is explained in Lemma 3.4 of \cite{Danupon-paths}.

For running time and edge congestion, \ref{BoundedDistanceSSSP} clearly runs in $K=(1+2/\e)h=O(\e^{-1}h)$ rounds, and every edge $(u,v)$ has at most two messages sent along it, one from $u$'s broadcast and one from $v$'s broadcast. \ref{BoundedHopSSSP} runs $O(\log n)$ iterations of \ref{BoundedDistanceSSSP}, so the total running time is $O(\e^{-1}h\log n)$ and the total edge congestion is $O(\log n)$.
\EP

\mylabel{BoundedHopSSSP}{\texttt{BoundedHopSSSP}}
\begin{algorithm}[H]
\caption{$\texttt{BoundedHopSSSP}(G=(V,E,w),s,h)$}
\small
\textbf{Input: }Weighted directed graph $G$, source vertex $s$, and integer $h$. $G$ has $s$-radius $\le n-1$.

\textbf{Output: } Every node $u$ knows the value of $\tilde d^h(s,u)$ such that $d^h(s,u)\le\tilde d^h(s,u)\le(1+\e)d^h(s,u)$.

\begin{algorithmic}[1]

\State Let $t$ be the time this algorithm starts. We can assume that all nodes know $t$.

\State\label{line:ForAllI}For all $i\in\lb\lc\log(n-1)\rc\rb$ and edge $(x,y)$, let $D_i'\gets2^i$ and $w_i'(x,y)\gets \lc\frac{2hw(x,y)}{\e D_i'}\rc$. Let $K\gets(1+2/\e)h$.

\For{\textbf{all} $i$}

\State $d'_i\gets \ref{BoundedDistanceSSSP}(G,w_i',s,K)$

\EndFor

        \State $\tilde d^h(s,u) \gets \f{\e D_i'}{2h}\min_id_i'(s,u)$

\end{algorithmic}
\end{algorithm}

\mylabel{BoundedDistanceSSSP}{\texttt{BoundedDistanceSSSP}}
\begin{algorithm}[H]
\small
\caption{$\texttt{BoundedDistanceSSSP}(G,w,s,K)$}
\textbf{Input: }Weighted directed graph $G$, source vertex $s$, and integer $K$.

\textbf{Output: } Every node $u$ knows $d'(s,u)$ where $d'(s,u)=d(s,u)$ if $d(s,u)\le K$ and $d'(s,u)=\infty$ otherwise.

\begin{algorithmic}[1]

\State Let $t$ be the time this algorithm starts. We can assume that all nodes know $t$.

\State Initially, every node $u$ sets $d'(s,u)\gets\infty$.

\State Source node $s$ sends a message $(s,0)$ to itself.

\If{a node $u$ receives a message $(s,\ell)$ for some $\el$ from node $v$}

  \If{$(\el+w(u,v)\le K)$ and $(\el+w(u,v)<d'(s,u))$}
    \State $u$ sets $d'(s,u)\gets\el+w(u,v)$.
  \EndIf

\EndIf

\State For any $x\le K$, at time $t+x$, every node $u$ such that $d'(s,u)=x$ broadcasts message $(s,x)$ to all its neighbors to announce that $d'(s,u)=x$.

\end{algorithmic}
\end{algorithm}

If we sample $k$ virtual nodes (including the source), we can run this algorithm with $h:=O(n\log n/k)$ from each virtual node, which is $O(k\log n)$ total congestion. Scheduling these algorithms in $O(\textsf{dilation}+\textsf{congestion}\cd\log n)$ time, we get $O(\e^{-1}h\logn+k\log^2n) = \tO(\e^{-1}n/k+k)$.

On the virtual graph, since all edge weights are multiples of $\f{\e D_i'}{2h}$, we can scale all weights by $\f{2h}{\e D_i'}$ so that they are integers, and then run \ref{Virtual} from \Cref{section:VirtualSSSP} in $O(kD\log n)$ time and $O(k\logn)$ congestion along each edge. Every node then scales its distance by $\f{\e D_i'}{2h}$, giving a $(1+\e)$-approximate distance from the source. Finally, extend the $(1+\e)$-approximate $h$-hop SSSP algorithm to all nodes in the same $\tO(\e^{-1}n/k+k)$ running time and $\tO(1)$ congestion, obtaining the following:

\BL\label{lemma:ApproximateSSSP}
We can compute $(1+\e)$-multiplicative approximate SSSP in $\tO(\e^{-1}n/k+kD)$ time and $\tO(k)$ congestion along each edge.
\EL

Applying \Cref{lemma:ApproximateSSSP} with $\e:=\al/n$ proves \Cref{lemma:AdditiveErrorSSSP}.

\section{Missing Details of \Cref{sec:main}}
\label{app:sec:main}

\mylabel{Main}{\texttt{Main}}
\begin{algorithm}[H]
\small
\caption{$\texttt{Main}(G,s)$}
\textbf{Input}: Every node $v$ knows the weight and direction of each edge $e\in E$ incident to $v$.

\textbf{Output}: Every node $v$ learns $d(v)$, which equals the exact distance $d_w(s,v)$.
\begin{algorithmic}[1]

\State $k\gets n^{3/4}D^{-3/4}$, $\el\gets n^{1/2}D^{1/2}$, $q\gets n^{1/4}D^{-1/4}$.


\For {$\lf\log_2\Lambda\rf+1$ iterations of \ref{Scaling}}
  \State $\tilde d \gets \el\texttt{-AdditiveErrorSSSP}(s,G,k)$
  \State Run $\ref{Small}(s,G,\tilde d,k,\el,q)$

\EndFor

\State Let $d$ be the distances output by \ref{Scaling}.

\end{algorithmic}
\end{algorithm}

\mylabel{SmallVirtual}{\texttt{SmallWeightVirtualSSSP}}
\begin{algorithm}
\small
\caption{$\texttt{SmallWeightVirtualSSSP}(s,G'=(V',E',w'))$}
\textbf{Input}: Every node $v'$ knows the weight and direction of each virtual edge $e'\in E'$ incident to $v'$. It is promised that $d_{w'}(s,t')\le |V'|-1$ for all nodes $t'$.

\textbf{Output}: Every node $v'$ learns $d(v')$, which equals the exact distance $d_{w'}(s,v')$.
\begin{algorithmic}[1]

\State $d(s)\gets 0$, $d(t')\gets \infty$ for all $t'\ne s$.

\State $i \gets 0$, $k\gets |V'|$

\While {$i \le k-1$}
  \State For each node $u'$, if $d(u')=i$ and $u'$ has not broadcasted yet, broadcast $d(u')$.
  \State If node $v'$ has edge $(u',v')$ and node $u'$ broadcasted, update $d(v')\gets \min\{d(v'),d(u')+w'(u',v')\}$
  \State If no node broadcasted, $i\gets i+1$
\EndWhile

\end{algorithmic}
\end{algorithm}

\section{Improved Virtual SSSP}\label{section:ImprovedVirtualSSSP}
In this section, we focus on the virtual SSSP problem, with the goal of developing faster virtual SSSP algorithms for large network diameters $D$, by reducing the dependency on $D$.

Let graph $G$ be the network graph with $n$ nodes. Assume that we are in a \ref{Scaling} iteration, so the graph $G$\ has $s$-radius at most $n-1$. Let $G'=(V',E',w')$ be the virtual graph with $n_{V'}$ nodes containing the source $s\in V'$. Note that $D$ is the diameter of the \textit{original} graph, which could be much larger than $n_{V'}$.
The running time $\tO(n_{V'}D)$ of \ref{Virtual}, presented in the previous section, is near-optimal for small $D$, i.e., $D=\tO(1)$. For larger $D$, the $\tO(D)$ factor increases the running time substantially. In this section, we present two alternative algorithms for computing SSSP on the virtual graph for larger values of $D$. Recall that our task is for every virtual node $v'\in V'$ to compute a value $d(v')$ that equals $d_{w'}(s,v')$.

\subsection{Algorithm 1: Gathering}\label{section:Gathering}

\mylabel{VirtualSSSPGather}{\texttt{VirtualSSSPGather}}
\paragraph{Algorithm \ref{VirtualSSSPGather}}
Note that the virtual graph $G'$ has size $O(n_{V'}^2)$, so one strategy is to broadcast the entire graph to all nodes in $O(n_{V'}^2+D)$ time using standard pipelining techniques; we call this procedure \textit{gathering} the graph. Then, all virtual nodes can locally compute their distance from $s$. We call this algorithm \ref{VirtualSSSPGather}.

\BL\label{thm:VirtualGather}
The irtual SSSP algorithm \ref{VirtualSSSPGather} runs in $O(n_{V'}^2+D)$ rounds.
\EL

\subsection{Algorithm 2: Virtualizing}\label{section:Virtualizing}

%

The virtual SSSP algorithm of this section mimics certain procedures in \ref{Main}, except modified to work on a virtual graph. To understand the modifications necessary, we first discuss how to transform generic distributed algorithms to work on a virtual graph.

\paragraph{Distributed Algorithms on Virtual Graph}
First, we explain how to schedule distributed algorithms on virtual graphs, following a natural extension to \Cref{section:Scheduling}.

For a moment, suppose that a virtual graph $G'$ is actually a network graph, in that nodes can communicate directly along edges in $G'$. Let $A_1,\ldots,A_k$ be distributed algorithms running independently on this  network graph $G'$. We place an additional restriction on each algorithm: on every round, a node can either \textit{broadcast} the same message to all of its neighbors, or not send anything. That is, a single node is not allowed to send different messages to different neighbors in a single round. Following \Cref{section:Scheduling}, let \textsf{dilation} denote the maximum running time of any algorithm $A_i$, and let \textsf{broadcasts} denote the maximum number of times a single node broadcasts, summed up over all $k$ algorithms. Formally, define $b_{v,i}$ as the number of times node $v$ broadcasts in algorithm $A_i$, and define $\textsf{broadcasts} := \max_v \sum_ib_{v,i}$. Now, revert back to the normal setting, where $G'$ is a virtual graph of a network graph $G$; as usual, we do not assume that edges of $G'$ belong to the network. The following claim relates the performances of $A_1,\ldots,A_k$ on the hypothetical network graph $G'$ to their performances on the actual network graph $G$.

\BCL\label{claim:VirtualAlgo}
 Let $A_1,\ldots,A_k$ be independent, distributed algorithms on $G'=(V',E')$ in which nodes are only allowed to broadcast. Suppose that $G'$ is a virtual graph of a network graph $G$ with diameter $D$. Then, we can run algorithms $A_1,\ldots,A_k$ in $G$ in time $O(\textsf{dilation} \cd D + \textsf{broadcasts} \cd n)$.
\ECL

\BP
 Let $b_{v,i,t} \in \{0,1\}$ denote whether or not node $v$ broadcasts on step $t$ of algorithm $A_i$. We can simulate round $t$ of each algorithm $A_i$ in $O(D+ \sum_{v,i} b_{v,i,t})$ steps by standard pipelining techniques. Since we need to simulate \textsf{dilation} rounds in total, the total running time is $O(\textsf{dilation}\cd D+\sum_{v,i,t}b_{v,t,i})$. The latter term is the total number of broadcasts over all nodes and all algorithms, which is at most $n\cd\textsf{broadcasts}$.
\EP

Whenever we transform a set of distributed algorithms to run on a virtual graph, we say we \defn{virtualize} the algorithms.

\subsubsection{Virtualizing Previous Tools}

Here, we discuss how to virtualize the scaling framework, $(1+\e)$-multiplicative approximate SSSP, and the \ref{Short} algorithm to work on virtual graphs. The generic process is the same for all three. First, we modify the algorithm so that, at each round, every node can either broadcast a message to all other nodes, or remain silent. Then, we replace each round $t$ of the original algorithm with a pipelined broadcast, which takes $O(b_t+D)$ time, where $b_t$ is the number of nodes that broadcast on round $t$. When virtualizing multiple algorithms in parallel, we invoke \Cref{claim:VirtualAlgo}.

\paragraph{Scaling Framework} To update edge weights on each iteration of \ref{Scaling}, it is enough for every virtual node to broadcast its computed distance, so that all nodes can recompute the weights of their edges. By standard pipelining techniques, this takes $O(n_{V'}+D)$ time per iteration. Since every virtual node has distance at most $n-1$ from source $s$, the scaling frameworks needs just $O(\log n)$ iterations, resulting in $\tO(n_{V'}+D)$ rounds spent reweighting edges.

\paragraph{ShortRange} Observe that, on each round of \ref{Short}, every node always communicates by sending the same message to its neighbors. Therefore, we can trivially modify the algorithm to run in $O(\el q+h)$ rounds with $O(h/q)$ broadcasts per node. If we run $k$ \ref{Short} algorithms in parallel, each node broadcasts $O(hk/q+k)$ times, so applying \Cref{claim:VirtualAlgo} gives the following.

\BL\label{lemma:ShortRangeVirtual}
We can simulate $k$ independent runs of the \ref{Short} algorithm on a virtual graph with $n_{V'}$ nodes in time $O((\el q+h)\cd D+(hk/q+k) \cd n_{V'})$.
\EL

\paragraph{$(1+\e)$-Approximate SSSP}
Consider the problem of computing $(1+\e)$-multiplicative approximate distances on a virtual graph, for some $\e\le 1$. The algorithm of \Cref{lemma:ApproximateSSSP} first runs $k$ independent copies of \ref{BoundedHopSSSP} for some parameter $k$, each of which can be modified to take $\tO(\e^{-1}n_{V'}/k)$ rounds with $\tO(1)$ broadcasts per node. The dilation is $\tO(\e^{-1}n_{V'}/k)$ and each node broadcasts $\tO(k)$ times total, so by \Cref{claim:VirtualAlgo}, everything can be pipelined to run in $\tO(\e^{-1}n_{V'}/k\cd D+kn_{V'})$ time. Next, the algorithm faces another virtual SSSP instance on $k$ nodes, which it solves using \ref{VirtualSSSPGather} in $O(k^2+D)$ time. Finally, the extension part of the algorithm also runs $k$ independent copies of \ref{BoundedHopSSSP}, and the analysis is identical. Thus, the total running time to compute $(1+\e)$-approximate distances from $s$ is \[\tO\lp \f{n_{V'}D}{\e k} +kn_{V'}+k^2+D \rp.\]
Setting $k:=\e^{-1/2}D^{1/2}$ gives the following. We can assume that $k\le n_{V'}$, since otherwise, the $kn_{V'}+D$ factor makes the running time at least $\Om(n_{V'}^2+D)$, and we are better off running \ref{VirtualSSSPGather} instead.

\BL\label{lemma:ApproximateVirtual}
We can compute $(1+\e)$-multiplicative approximate distances on a virtual graph with $n_{V'}$ nodes in time $\tO ( \e^{-1/2}n_{V'}D^{1/2} )$.
\EL

\subsubsection{Warm-up: Nonrecursive Virtual SSSP}

Here, we describe the virtual SSSP algorithm \ref{VirtualSSSPNonrecursive} that uses the virtualized scaling framework and \ref{Short} algorithm. It runs in $\tO(n_{V'}r^{1/2}D^{1/2}+D)$ time on any virtual graph with $s$-radius at most $r$. Recall that, after running \ref{Scaling}, $r$ can be as large as $n_{V'}-1$, giving a running time of $\tO(n_{V'}^{3/2}D^{1/2}+D)$. Then, in \Cref{sec:Recursive}, we will make this algorithm recursive and further improve the running time.

\paragraph{Virtualized Scaling Framework}
The algorithm first applies the virtualized \ref{Scaling} to the virtual graph $G'$ on $n_{V'}$ nodes, reducing the task to computing $O(\log n)$ iterations of virtual SSSP on a graph with $s$-radius at most $n_{V'}-1$. As discussed before, the total time spent reweighting edges in the virtualized \ref{Scaling}  is $\tO(n_{V'}+D)$.

\paragraph{Virtualized ShortRange}
The algorithm first samples a set of nodes $V''\s V'$, where every virtual node in $V'$ is sampled with probability $k/n_{V'}$, for some parameter $k$, except that $s\in V''$ with probability $1$. We run the virtualized \ref{Short} with parameters $h:=O(n_{V'}\log n_{V'}/k)$ and $\el:=r$ from each node in $V''$, taking $O((\el q+h)\cd D+(hk/q+k) \cd n_{V'})$ time with a free parameter $q\ge1$, by \Cref{lemma:ShortRangeVirtual}. 

 Then, following notation of line~\ref{line:BuildGraph} of \ref{Small}, we build a graph $G''=(V'',E'')$, where for each pair $(u'',v'')\in V''\times V''$ with $d_{u''}(v'')<\infty$, add to $E''$ a directed edge $(u'',v'')$ of weight $d_{u''}(v'')$ that is known to node $v''$. We compute distances in $G''$ by gathering the graph in $O(k^2+D)$ time, so that each node in $V''$ can locally compute its correct distance. Finally, all nodes in $V''$ broadcast their distances, allowing each remaining node $v'\in V'$ to compute its distance $d(v')=\min_{u''\in V''} (d(u'')+d_{u''}(v'))$. 
We next show that, w.h.p., $d(v')$ is the correct distance for each $v'\in V'$.

Consider one iteration of \ref{Scaling}, and fix a node $t'\in V'$ in the above algorithm. By \Cref{lemma:Virtual}, there is a  shortest $s\to t'$ path $P$ in $G'$ with a node in $V''$ within every consecutive $h$ nodes on the path. In other words, consecutive nodes in $V''\cap P$ are within $h$ hops from each other on the path. Since $G'$ has $s$-radius at most $r$, we have $w'(P)\le r$, so these consecutive nodes are also within distance $r=\el$. Hence, by the guarantees of \ref{Short}, for every two consecutive nodes $u'',v''\in V''$, $d_{u''}(v'')$ is the correct distance $d_{w'}(u'',v'')$, and there is an edge $(u'',v'')\in E''$ with weight $d_{w'}(u'',v'')$. Therefore, after computing SSSP on $G''$, all nodes in $V''\cap P$ know their correct distance. In particular, let $t''$ be the last node on the path that is in $V''$. Since $t$ is within $h$ hops and distance $\el$ from $t''$ along $P$, we also have $d_{t''}(t')=d_{w'}(t'',t')$, so \[d_{w'}(s,t') = d_{w'}(s,t'')+d_{w'}(t'',t)=d(t'')+d_{t''}(t').\]
 Since node $t'$ computes $d(t')=\min_{u''\in V''} (d(u'')+d_{u''}(t'))$ and the distances $d(v')$ can never be a strict underestimate, we have $d(t')=d_{w'}(s,t')$, as desired.

Finally, we analyze the running time. Overall, the total number of rounds is \[ O\lp (\el q+h)\cd D+\f{hkn_{V'}}q  + kn_{V'}+k^2+D \rp = \tO\lp r qD+\f{n_{V'}D}{k}+\f{n_{V'}^2}q +kn_{V'}+D \rp, \]
where we recall that $\el=r$, $h=\tO(n_{V'}/k)$, and $ kn_{V'} \ge k^2$. Setting $k:=D^{1/2}$ and $q:=n_{V'}r^{-1/2}D^{-1/2}$ gives running time $\tO(n_{V'}r^{1/2}D^{1/2}+D)$, as desired.  Note that, in the event that $k>n_{V'}$, we have $D > n_{V'}^2$, so we are better off running  \ref{VirtualSSSPGather} on $G'$ in the optimal $O(n_{V'}^2+D)=O(D)$ time. Also, if  $q<1$, the $n_{V'}^2/q$ factor becomes at least $n_{V'}^2$ and again, we are better off running \ref{VirtualSSSPGather}. Thus, we can ignore these two corner cases.

Therefore, we have proved the following.

\BL\label{lemma:VirtualSSSP54}
The algorithm \ref{VirtualSSSPNonrecursive} computes exact distances on a virtual graph with $n_{V'}$ nodes and $s$-radius at most $r$ in time $\tO(n_{V'}r^{1/2}D^{1/2}+D)$.
\EL

\mylabel{VirtualSSSPNonrecursive}{\texttt{VirtualSSSPNonrecursive}}
\begin{algorithm}[H]
\small
\caption{$\ref{VirtualSSSPNonrecursive}(s,G'=(V',E',w'),k,q)$}
\textbf{Input}: Every node $v'\in V'$ knows the weight and direction of every edge $(u',v')$ in $E'$. We have the promise that $d_{w'}(s,t')\le r,\forall t'\in V'$.

\textbf{Output}: Every node $v$ learns $d(v)$, which equals the exact distance $d_w(s,v)$.
\begin{algorithmic}[1]

\State Sample a set of nodes $V''\s V'$. Every node joins the set with probability $k/n$, except for $s$, which joins with probability $1$.

\State Set $h\gets\Th(n_{V'}\log n_{V'}/k)$ and $\el\gets r$. For each $s''\in V''$, let $d_{s''}$ be the result of running a virtualized $\ref{Short}(s'',G,h,\el,q)$


\State Build a graph $G''=(V'',E'')$, where for each pair $(u'',v'')\in V''\times V''$ with $d_{u''}(v'')<\infty$, add to $E''$ a directed edge $(u'',v'')$ of weight $d_{u''}(v'')$ that is known to node $v''$. 

\State Run \ref{VirtualSSSPGather} on the virtual graph $G''$, so that each node $v''\in V''$ knows its distance $d(v'')$ in $G''$.

\State Every node $u''\in V''$ broadcasts its distance $d(u'')$ to all other nodes, and every node  $v'\in V'\setminus V''$ computes $d(v')\gets\min_{u''\in V''} (d(u'')+d_{u''}(v'))$

\end{algorithmic}
\end{algorithm}

\subsubsection{Recursive Virtual SSSP}\label{sec:Recursive}

In this section, we extend \ref{VirtualSSSPNonrecursive} to work recursively, arriving at our final algorithm \ref{VirtualSSSPRecursive}. Our goal is to compute virtual SSSP in $\tO(n_{V'}r^\e D^{1/2} + r^{1-2\e}D)$ time for any constant $\e\in(0, 1/2]$, on a graph with $s$-radius at most $r$. Note that the case $\e=1/2$ is covered by \ref{VirtualSSSPNonrecursive} and \Cref{lemma:VirtualSSSP54}.

\paragraph{Recursion}
For ease of notation, define a virtual SSSP algorithm to be \textit{$\e$-good} if, for any integers $n_{V'}$ and $r$, and on any virtual graph with $n_{V'}$ nodes and $s$-radius $r$, the algorithm runs in $\tO(n_{V'}r^\e D^{1/2} + r^{1-2\e}D)$ time. Our $\e$-good algorithm will make a recursive call to an $\e'$-good algorithm, for some $\e'>\e$. Eventually, the value of $\e$ in the recursion goes up to $1/2$, at which point the algorithm of \Cref{lemma:VirtualSSSP54} serves as the base case and performs the job.

\paragraph{Additive Approximation}
Let $G'$ be a virtual graph with $n_{V'}$ nodes and $s$-radius at most $r$. The algorithm begins by computing an $\el$-additive approximate SSSP, for some $\el\le r$. Note that, since the $s$-radius is bounded by $r$ this time, a $(1+\e)$-multiplicative approximation for $\e:=\el/r$ suffices. By \Cref{lemma:ApproximateVirtual}, this step has round complexity of \[\tO \lp \f{n_{V'}D^{1/2}}{\e^{1/2}} \rp = \tO\lp \f { n_{V'}r^{1/2}D^{1/2}}{\el^{1/2}}\rp. \]

\paragraph{Bucketing}
Similarly to \ref{Small}, the algorithm defines buckets so that bucket $V'_i$ contains all virtual nodes whose distance from $s$ is in the range $[i\el,(i+1)\el)$. The key difference is that, since the $s$-radius is at most $r$ this time, we only need $O(r/\el)$ buckets.
Then, use the $\el$-additive approximation $\tilde d$ to group each virtual node $v'\in V'$ into at most two buckets, according to the rule $v'\in V'_i \iff \tilde d(s,v') \in [i\el,(i+2)\el)$.  From now on, let $n_i:=|V'_i|$ be the number of nodes in bucket $n_i$, so that $\sum_in_i\le 2n_{V'}$.

\paragraph{Graph on Each Bucket}
The algorithm then processes the buckets $V'_i$ in increasing order of $i$. For each bucket $V'_i$, we assume the following invariant right before processing bucket $V'_i$, which is satisfied vacuously for the first iteration $i=0$:

\medskip
\begin{itemize}
\item \textbf{(Invariant)}: Right before we process bucket $V'_i$, all virtual nodes whose distance from $s$ is less than $i\el$ has broadcast its correct distance to all other virtual nodes.
\end{itemize}
\medskip

\noindent Every virtual node $v'\in V'_i$ first computes its best distance $d_i(v')$ based on distance broadcasts from its neighbors in $G'$ on iterations before $i$. Now build a virtual graph $\tilde G_i'=(\tilde V'_i,\tilde E'_i)$ as follows: take the subgraph of $G'$ induced by $V'_i$, add $s$ to the graph, and direct an edge from $s$ to each virtual node $v'\in V'_i$ with edge weight $d_i(v')-(i-1)\el$. Observe that all edge weights are nonnegative, since $d_i(v') \ge d_{G'}(s,v') \ge \tilde d_{G'}(s,v')-\el\ge i\el-\el$. We now claim the following:

\BC\label{claim:GraphBucket}
For all virtual nodes $t'\in V'_i$, $d_{\tilde G_i'}(s,t')\ge d_{G'}(s,t')-(i-1)\el$. If node $t'$ also satisfies $d_{G'}(s,t')\in[i\el,(i+1)\el)$, then we have $d_{\tilde G_i'}(s,t')=d_{G'}(s,t')-(i-1)\el$.
\EC
\BP
It is easy to see that $d_{\tilde G_i'}(s,t')\ge d_{G'}(s,t')-(i-1)\el$, since edges do not get shorter except for the added edges from $s$, so we focus on the other direction.

Fix a virtual node $t'\in V'_i$. We know that $d_{G'}(s,t')\le \tilde d(s,t') < (i+2)\el$. Let $P$ be the shortest $s\to t'$ path in $G'$, and let $u'$ be the last virtual node on $P$ satisfying $d_{G'}(s,u')<i\el$. Then, $u'$ has already broadcast its correct distance, and its next node $v'$ on $P$ has computed its correct distance $d_{G'}(s,v')$ in $G'$. In particular, there is an edge from $s$ to $v'$ with weight $d_{G'}(s,v')-(i-1)\el$ in $\tilde G'_i$. Also, all nodes from $v'$ to $t'$ on $P$ are in $V'_i$. Therefore, the subpath of $P$ from $v'$ to $t'$, with $s$ appended at the front, has length exactly $d_{G'}(s,t')-(i-1)\el$ in $\tilde G'_i$. This proves $d_{\tilde G_i'}(s,t')\le d_{G'}(s,t')-(i-1)\el$.
\EP

In particular, all virtual nodes $v'\in V$ satisfying $d_{G'}(s,v')\in[i\el,(i+1)\el)$ have distance at most $2\el$ from $s$ in $\tilde G_i'$. Next, the algorithm computes the correct distance from $s$ to these nodes in $\tilde G_i'$ through a recursive call.

\paragraph{Recursive SSSP on Each Bucket}
The $\e$-good algorithm computes a virtual SSSP on $\tilde G_i$ recursively by calling an $\e'$-good algorithm, for some $\e'>\e$; the relation between $\e'$ and $\e$ will be determined later. $\tilde G_i$ is a graph with $n_i$ nodes and $s$-radius at most $2\el$, so the recursive call takes \[\tO(n_i\el^{\e'} D^{1/2} + \el^{1-2\e'}D)\] time. After the recursive call, each virtual node $v'\in V'_i$ takes its computed distance $\tilde d_i(v')$ and adds $(i-1)\el$ to it, and broadcasts as its distance estimate for $d_{G'}(s,v')$ to all other nodes. In particular, all virtual nodes $v'\in V'$ with $d_{G'}(s,v')\in[i\el,(i+1)\el)$ broadcast their correct distance, satisfying the Invariant. Note that broadcasting the distances takes $n_i+D$ time, which does not affect the running time asymptotics.

\paragraph{Running Time}
Summing up the running time of the approximate SSSP and the recursive virtual SSSP calls in each bucket, we obtain: \[\tO \lp \f{n_{V'}r^{1/2}D^{1/2}}{\el^{1/2}} \rp + \sum_{i=1}^{O(r/\el)} \tO\lp n_i\el^{\e'} D^{1/2} + \el^{1-2\e'}D \rp = \tO\lp \f{n_{V'}r^{1/2}D^{1/2}}{\el^{1/2}} + n_{V'}\el^{\e'}D^{1/2} + \f{rD}{\el^{2\e'}}\rp, \]
using the fact that $\sum_in_i\le 2n_{V'}$. Setting $\el:=r^{1/(1+2\e')}$ gives running time \[ \tO(n_{V'}r^{\e'/(1+2\e')} D^{1/2} + r^{1-2\e'/(1+2\e')}D).\] In other words, the algorithm is $\e$-good for $\e:=\e'/(1+2\e')$.

Moreover, setting $\el:=r^{(1+\de)/(1+2\e')}$ for $\de < 2\e'$ gives running time \[\tO(n_{V'}r^{(1-(1+\de)/(1+2\e'))/2} D^{1/2} + r^{1-2\e'(1+\de)/(1+2\e')}D) ,\]
and the exponent of $r$ in the first term satisfies $(1-(1+\de)/(1+2\e'))/2 \le \e'(1+\de)/(1+2\e')$, so the algorithm is $\e'(1+\de)/(1+2\e')$-good. Therefore, by choosing an appropriate $\de$, the algorithm can become $\e$-good for any $\e$ satisfying \[\e'/(1+2\e') \le \e < \e' .\]

\paragraph{Solving the Recursion}
Finally, we prove the desired running time of the algorithm.

\BL\label{lemma:Recursion}
For any constant $\e\in(0,1/2]$, there is a virtual SSSP algorithm running in $\tO(n_{V'}r^\e D^{1/2} + r^{1-2\e}D)$ rounds on a virtual graph with $n_{V'}$ nodes and $s$-radius at most $r$.\EL
\BP
We prove by induction on $\al\in\N$ that there is an $\e$-good virtual SSSP algorithm for all $\e \in [1/(2\al), 1/2]$. The base case $\al=1$ follows by \Cref{lemma:VirtualSSSP54}.

Now assume that there is an $\e'$-good virtual SSSP algorithm for all $\e'\in[1/(2\al),1/2]$. In particular, the statement is true for $\e'=1/(2\al)$. By our previous observation, in one recursive step, we obtain an $\e$-good algorithm for any $\e$ satisfying $\e'/(1+2\e') \le \e<\e'$, which comes out to $\e \in [1/(2\al+2), 1/(2\al))$. Hence, there is an $\e$-good virtual SSSP algorithm for all $\e\in[1/(2\al+2),1/2]$, completing the induction.
\EP

Observe that the above claim only holds for constant values of $\e$, since the number of virtual nodes can double upon a recursive call. That is, we need to maintain a constant number of recursion levels.

\paragraph{Final Running Time}
Finally, due to the virtualized \ref{Scaling}, we can assume that $r\le n_{V'}-1$ initially, giving our main result in this section.

\BT\label{thm:VirtualSSSPEps}
For any constant $\e \in (0, 1/2]$, there is a virtual SSSP\ algorithm with round complexity of $\tO(n_{V'}^{1+\e}  D^{1/2} + n_{V'}^{1-2\e}D)$.
\ET

\mylabel{VirtualSSSPRecursive}{\texttt{VirtualSSSPRecursive}}
\begin{algorithm}[H]
\small
\caption{$\ref{VirtualSSSPRecursive}(s,G'=(V',E',w'),\e)$}
\textbf{Input}: Every node $v'\in V'$ knows the weight and direction of every edge $(u',v')$ in $E'$. We have the promise that $d_{w'}(s,t')\le r,\forall t'\in V'$.

\textbf{Output}: Every node $v$ learns $d(v)$, which equals the exact distance $d_w(s,v)$.
\begin{algorithmic}[1]

\State Let $\e'$ be the solution to $\e=\e'/(1+2\e')$, or $\e'\gets1/2$ if the solution is greater than $1/2$.

\State Let $\de$ be the solution to $\e=\e'(1+\de)/(1+2\e')$.

\State $\el \gets r^{(1+\de)/(1+2\e')}$, $k \gets (\el/r)^{-1/2}D^{1/2}$

\State $\tilde d \gets \el\texttt{-AdditiveErrorVirtualSSSP}(s,G',k)$

\State Define $V'_i$ as $\{ v'\in V' : \tilde d(s,v')\in[i\el,(i+2)\el) \}$.

\State
  Initialize distances $d(s)\gets0$, $d(v)\gets\infty$ for all $v\in V\setminus s$.


\For {$i=0,\ldots,\lf r/\el\rf$} {\tt\color{gray} // Iterate over buckets $V_0,\ldots,V_{\lf r/\el\rf}$.}
  
  \State
    Build the graph $\tilde G'_i=(\tilde V'_i,\tilde E'_i)$ as follows. Start with the graph $G'$ induced by $V'_i$. Then, add node $s$ to $\tilde V'_i$ if necessary, and for each $v'\in V'_i$, add to $\tilde E'_i$ a directed edge $(s,v')$ of weight $d(v')-(i-1)\el$ that is known to node $v'$.
  \If {$\e'=1/2$} {\tt\color{gray} // Base case.}
    \State $k\gets D^{1/2}$, $q\gets |\tilde V'_i|r^{-1/2}D^{-1/2}$
    \State
      Run $\ref{VirtualSSSPNonrecursive}(s,\tilde G_i',k,q)$, and let $\tilde d_i'$ be the resulting distance that each node in $\tilde V'_i$ learns.
  \Else {\tt\color{gray}\ // Recursive step.}
    \State
      Run $\ref{VirtualSSSPRecursive}(s,\tilde G_i',\e')$, and let $\tilde d_i'$ be the resulting distance that each node in $\tilde V'_i$ learns.
  \EndIf
  \State
    Every node $v'\in V'_i$ broadcasts $\tilde d'_i(v')$ to all other virtual nodes. {\tt\color{gray}// These broadcasts are pipelined along a BFS tree.}
  \State
    Every node $v'\in V'$ sets $d(v')=\min\{d(v'),d(u')+w'(u',v')\}$ for each $\tilde d_i(u')$ received from neighbor $u'$ in $G'$.

\EndFor

\end{algorithmic}
\end{algorithm}

\subsection{Running Time Improvements}

From the proof of \Cref{lemma:SSSPRunningTime}, observe that
\ref{Small} runs in \[\tO\lp \el q+\f nq+\f{n^2}{\el k}\rp\] time, plus the time it takes to solve a virtual SSSP instance on $k$ virtual nodes. If we use \ref{VirtualSSSPGather} of \Cref{thm:VirtualGather} that runs in $O(k^2+D)$ rounds, we get running time \[ \tO\lp \el q+\f nq+\f{n^2}{\el k}+k^2+D\rp ;\] setting $k:=n^{3/7}$, $\el:=n^{5/7}$, and $q:=n^{1/7}$ gives total running time $\tO(n^{6/7}+D)$.

If we use the algorithm of \Cref{thm:VirtualSSSPEps}
that runs in $\tO(k^{1+\e}  D^{1/2} + k^{1-2\e}D)$  rounds, we get running time  \[ \tO\lp \el q+\f nq+\f{n^2}{\el k}+k^{1+\e}  D^{1/2} + k^{1-2\e}D\rp, \] as long as $\e \in [\Om(1), 1/2]$. To optimize this expression,  we first set $\e$ to be the solution to the equation $k^{3\e}=D^{1/2}$ so that, for any value of $k$, \[ k^{1+\e}  D^{1/2} + k^{1-2\e}D = k D^{2/3}. \]
Then, set $k:=n^{3/4}D^{-1/2}$, $\el:=n^{1/2}D^{1/3}$, and $q:=n^{1/4}D^{-1/6}$ to obtain running time $\tO(n^{3/4}D^{1/6})$, as long as $\e\in[\Om(1),1/2]$.

We first restrict ourselves to the case $D=n^{\Om(1)}$, which implies $\e=\Om(1)$. As for the constraint
 $\e\le1/2$, that happens precisely when \[ D^{1/2}=k^{3\e} \le k^{3/2} = (n^{3/4}D^{-1/2})^{3/2} \iff  D\le n^{9/10} . \]
 
 Thus, we have an $\tO(n^{3/4}D^{1/6})$ algorithm when $n^{\Om(1)} \le D\le n^{9/10}$, and an $\tO(n^{6/7}+D)$ algorithm for any $D$. Note that when $D \ge n^{6/7}$, we are always better off running the $\tO(n^{6/7}+D)$ algorithm, so the upper bound of $D\le n^{9/10}$ for the virtualizing SSSP case does not hinder us overall. Hence, the optimal algorithm of the two has a running time expressible as \[\tO \lp \min\{ n^{3/4}D^{1/6},n^{6/7}\} + D\rp .\] 
For the case when  $D=n^{o(1)}$, the $\tO(n^{3/4}D^{1/4})$ algorithm of \Cref{thm:main1} runs in $\tO(n^{3/4+o(1)})$ time. Altogether, by choosing the best algorithm of the three, we obtain  the running time \[\tO\lp n^{3/4+o(1)}+ \min\{ n^{3/4}D^{1/6},n^{6/7}\}+D\rp ,\]
 proving \Cref{thm:main2}.

\section{Multiple sources}\label{section:MultipleSources}

Consider now the multiple source SSSP problem: given $\kappa$ sources, we want each node $v\in V$ to know its exact distance from each source. In this section, we modify our single-source SSSP algorithm to obtain efficient algorithms for the $\kappa$-sources problem.

Our high-level ideas are the same: we run the scaling framework $O(\log\Lambda)$ times, and we solve each iteration for all $\kappa$ sources in parallel. To elaborate, we first construct $\kappa$ graphs $G_s$, one for each source $s$, so that every graph $G_s$ has $s$-radius at most $n-1$. In $\kappa$ rounds, every node can know its incident edges in each graph $G_s$. We then run $\ref{Small}(s,G_s,k,\el,q)$ from each source $s$ \textit{in parallel}, once again using \cite{Scheduling} to schedule our algorithms, so that the entire algorithm runs in $\tO((\textsf{dilation}+\kappa\cd\textsf{congestion})\cd \log \Lambda)$ time. where \textsf{dilation} and \textsf{congestion} are parameters of $\ref{Small}(s,G_s,k,\el,q)$ for a single source $s$.

Since we have multiple virtual SSSP algorithms, we will break the running time inside and outside the virtual SSSP computation.

\paragraph{Outside Virtual SSSP}
The dilation $\tO\lp \el q+\f nq+\f{n^2}{\el k}\rp$ of \ref{Small} outside of the virtual SSSP algorithm follows from \Cref{lemma:SSSPRunningTime}. The following lemma analyzes the congestion.

\BL\label{SSSPCongestion}
In one execution of \ref{Small} outside of virtual SSSP, $\tO(n/q +k+n/k )$ messages are passed along each edge.
\EL
\BP
By \Cref{lemma:AdditiveErrorSSSP}, $\el\texttt{-AdditiveErrorSSSP}$ has congestion $\tO(k)$. Again, we use the same value of $k$ for $\el\texttt{-AdditiveErrorSSSP}$.
Also,
\ref{Short} from all virtual nodes takes $\tO(n/q)$ congestion, following the proof of \Cref{lemma:SSSPRunningTime}.

We now argue about the congestion of \ref{Extend}. For an edge $(u,v)$, let $i$ be the first iteration when $d(u)$ is sent from $u$ to $v$. It follows that the current estimate $d(u)$ satisfies $d(u)+w(u,v)\in[i\el,(i+1)\el)$. Since estimates only become lower over time, we still have $d(u)<(i+1)\el$ at the end of iteration $i$. Therefore, no message is sent from $u$ to $v$ in future iterations of \ref{Extend}. It follows that $u$ only sends messages to $v$ in one iteration, which is $O(h)$ messages. We can repeat the same argument to bound the number of messages from $v$ to $u$, so the congestion along each edge is $O(h)=\tO(n/k)$.
\EP

Hence, by the $\tO(\textsf{dilation}+\kappa\cd\textsf{congestion})$ formula, the total running time of \ref{Small} outside virtual SSSP is \[ \tO \lp \el q+\f nq+\f{n^2}{\el k} + \kappa \lp \f nq+k+\f nk \rp \rp .\]

\paragraph{Inside Virtual SSSP}
Now we consider the changes to virtual SSSP algorithms. Again, let $n_{V'}$ be the number of nodes in each of the $\kappa$ virtual graphs.

First, consider the original \ref{Virtual} algorithm of \Cref{lemma:VirtualSSSP}. In \ref{Virtual}, the total number of broadcasts is $\tO(\kappa n_{V'})$ over all $\kappa$ algorithms, so the congestion is $\tO(n_{V'})$. Adding this onto the dilation, we obtain running time \[ \tO(\kappa n_{V'} + n_{V'}D) . \]


Now consider the two improved algorithms of \Cref{section:ImprovedVirtualSSSP}. The gathering algorithm gathers $\kappa$ graphs of size $O(n_{V'}^2)$, so it takes \[O(\kappa n_{V'}^2+D)\] rounds total.

For the virtualizing algorithm, we first modify the virtualized procedures to run $\kappa$ iterations in parallel. The only difference in the running time is the value of \textsf{broadcasts}, which increases by a factor of $\kappa$. We again use \Cref{claim:VirtualAlgo} and optimize the parameters.
\BE
\im Reweighting edges in the scaling framework now takes $\tO(\kappa n_{V'}+D)$ rounds.
\im The virtualized $(1+\e)$-multiplicative approximate SSSP now takes $\tO(n_{V'}D/(\e k) + \kappa kn_{V'} + k^2+D)$ rounds. Setting $k:=\kappa^{-1/2}\e^{-1/2}D^{1/2}$ gives running time $\tO(\kappa^{1/2}\e^{-1/2}n_{V'}D^{1/2})$. Here, to ensure that $k \ge 1$, we impose the restriction $\kappa < D$. Also, if $k>n_{V'}$, then the $\kappa kn_{V'}+D$ factor makes the running time at least $\Om(\kappa n_{V'}^2+D)$, and we are better off gathering the graph.
\im The virtualized \ref{Short} now takes $\tO(rqD + n_{V'}D/k + \kappa n_{V'}^2/q + \kappa n_{V'}k +D)$ rounds. We can optimize $k:= \kappa^{-1/2}D^{1/2}$ and $q:=\kappa^{1/2}n_{V'}r^{-1/2}D^{-1/2}$ to get running time $\tO(\kappa^{1/2} n_{V'}r^{1/2}D^{1/2}+D )$. Again, we assume that $\kappa < D$. Also, if $q<1$, then the $\kappa n_{V'}^2/q+D$ factor is at least $\Om(\kappa n_{V'}^2+D)$.
\EE

That is, both running times increase by a factor of $\kappa^{1/2}$. We can work through the recursion in \Cref{lemma:Recursion} with these updated values in the same manner and obtain the following.

\BL
For any $\e\in(0,1/2]$, there is an algorithm that computes virtual SSSP on $\kappa$ graphs in parallel, each with $n_{V'}$ nodes and $s$-radius at most $r$. It runs in $\tO(\kappa^{1/2}n_{V'}r^\e D^{1/2} + r^{1-2\e}D)$ rounds.
\EL

Since $r=n_{V'}-1$ by the scaling framework, the virtualizing algorithm solves $\kappa$ virtual SSSPs in \[\tO(\kappa^{1/2} n_{V'}^{1+\e}D^{1/2} + n_{V'}^{1-2\e}D)\] rounds.

Lastly, we optimize for the parameters in each of the virtual SSSP algorithms. We have three cases, some of which impose constraints on $\kappa$ and $D$.

\BE
\im $\kappa \ge Dn^{-o(1)}$. In this case, we run the original \ref{Virtual} in $\tO(\kappa k +kD) = \tO(\kappa k)$ rounds, for a total running time of \[ \tO \lp \el q+\f nq+\f{n^2}{\el k} + \kappa \lp \f nq+k+\f nk \rp + kD \rp . \]
Setting $k:=n^{3/4}\kappa^{-1/2}$, $\el:=n^{1/2}$, and $q:=n^{1/2}\kappa^{-1/2}$ gives a final time of \[\tO(\kappa^{1/2} n^{3/4+o(1)}).\]

\im The gathering virtual SSSP algorithm in $\tO(\kappa k^2+D)$ rounds, for a total running time of \[ \tO \lp \el q+\f nq+\f{n^2}{\el k} + \kappa \lp \f nq+k+\f nk \rp + \kappa k^2+D \rp .\]
Setting $k:=\kappa^{-2/7}n^{3/7}$, $\el:=\kappa^{-1/2}n^{5/7}$, and $q:=\kappa^{4/7}n^{1/7}$ gives a final time of \[\tO(\kappa^{3/7}n^{6/7}+D).\]

\im $\kappa \le Dn^{-\Om(1)}$ and $D \le \kappa^{2/5}n^{9/10}$. In this case, we use the virtualizing algorithm in $\tO(\kappa^{1/2} k^{1+\e}D^{1/2} + k^{1-2\e}D)$ rounds. We first set $\e$ to satisfy $\kappa^{1/2} k^{1+\e}D^{1/2} = k^{1-2\e}D \iff k^{3\e}=\kappa^{-1/2}D^{1/2} $, so that the algorithm runs in $\tO(\kappa^{1/3}n_{V'}D^{2/3})$ rounds.
Note that $\e \in [\Om(1), 1/2]$ in this case. The total running time becomes \[ \tO \lp \el q+\f nq+\f{n^2}{\el k} + \kappa \lp \f nq+k+\f nk \rp + \kappa^{1/3}kD^{2/3} \rp  .\]
Setting $k:=n^{3/4}D^{-1/2}$, $\el:=\kappa^{-1/3}n^{1/2}D^{1/3}$, and $q:=\kappa^{2/3}n^{1/4}D^{-1/6}$ gives a final time of \[ \tO(\kappa^{1/3}n^{3/4}D^{1/6}) .\]
We now verify the constraints on $\e$. Since $k^{3\e} = \kappa^{-1/2}D^{1/2}=n^{\Om(1)}$ by assumption, we have $\e=\Om(1)$. Also,
\[\e\le1/2\iff  \kappa^{-1/2}D^{1/2}=k^{3\e} \le k^{3/2}=(n^{3/4}D^{-1/2})^{3/2}\iff D\le \kappa^{2/5} n^{9/10} , \]
which is precisely our constraint.

Note that when $D>\kappa^{2/5}n^{9/10}$, we also have $D > \kappa^{3/7}n^{6/7}$, so we are better off running the gathering virtual SSSP instead, giving an optimal $\tO(\kappa^{3/7}n^{6/7}+D)=\tO(D)$ rounds total. Therefore, the $D\le\kappa^{2/5}n^{9/10}$ restriction does not affect us overall.
\EE

Altogether, the three cases combined produce an algorithm that runs in time \[ \tO \lp \min \left\{ \kappa^{1/2}n^{3/4+o(1)}+ \kappa^{1/3}n^{3/4}D^{1/6} , \kappa^{3/7}n^{6/7} \right\} + D \rp ,\] proving \Cref{thm:main3}.

\end{document}